%% file: main.tex
\documentclass[showpacs,amsmath,amssymb]{revtex4}
\usepackage[english]{babel}
\usepackage[utf8]{inputenc}
\usepackage{amsmath,amssymb}
\usepackage{graphicx}
\usepackage{hyperref}
\usepackage{wrapfig,framed}
\usepackage{ulem, cancel} 
\usepackage{url}
\usepackage{longtable} 
\usepackage{array}
\usepackage[colorinlistoftodos]{todonotes}
\usepackage{tikzfeynman}
\usepackage{adjustbox}

\newcommand{\be}{\begin{equation}}
\newcommand{\ee}{\end{equation}}

\newcommand{\gStar}{g_{\star}}
\newcommand{\sHat}{\hat{S}}
 
\begin{document}
\title{Critical exponent $\eta$ in 2D $O(N)$-symmetric $\varphi^4$-model up to 6~loops}
\author{L.~Ts. \surname{Adzhemyan}}
\author{ Yu.~V. \surname{Kirienko}}
\author{M.~V. \surname{Kompaniets}}
\email{m.kompaniets@spbu.ru}

\affiliation{%
St. Petersburg State University, 
7/9 Universitetskaya nab., 
St. Petersburg, 199034 
Russia.%
}
\date{\today}

\begin{abstract}
Critical exponent $\eta$ (Fisher exponent) in $O(N)$-symmetric $\varphi^4$-model was calculated  using renormalization group approach in the space of fixed dimension $D=2$ up to 6~loops.
The calculation of the renormalization constants was performed with the use of $R'$-operation and specific values for diagrams were calculated in Feynman representation using sector decomposition method. Presented approach allows easy automation and generalization for the case of complex symmetries. Also a summation of the perturbation series was obtained by Borel transformation with conformal mapping.
The contribution of the 6-th term of the series led to the increase  of the Fisher exponent in $O(1)$ model up to $8\%$.
\end{abstract}
\maketitle

\section{Introduction}
Currently, the renormalization group (RG) method is a common tool in the theory of critical behavior and phase transitions. It enables us to justify the critical scaling and to calculate the critical exponents and universal amplitude ratios in the form of «regular» perturbation theory.
Such a theory can be constructed in the framework of $\varepsilon$-expansion~\cite{Vasilev2004, Zinnbook}; another popular approach is the renormalization group in the fixed space dimension~\cite{Nickel1977, Nickel1978,Orlov2000,Calabrese}.

One of the powerful tools used in the framework of the $\varepsilon$-expansion is sector decomposition technique~\cite{Heinrich2008}. This technique allows one to efficiently allocate poles in the Feynman diagrams and to represent the residues at the poles in a form of well convergent integrals of functions with bounded variation which can be calculated numerically. Despite the fact that this technique was initially developed for extracting poles in $\varepsilon$ from diagrams, it is extremely powerful in computing convergent Feynman integrals and allows one to calculate these integrals numerically with high precision.

In papers~\cite{Adzhemyan2013representation,Adzhemyan2013five-loop} the approach that allows to construct $\varepsilon$-expansion of critical exponents in terms of convergent integrals («theory without divergences»~\cite{zavialov,Adzhemyan2013representation}) was proposed. It turns out that for such integrals sector decomposition is very useful and allows to calculate them with high accuracy. Combining «theory without divergences» and sector decomposition~\cite{Adzhemyan2013five-loop} allowed to calculate critical exponents of the $\varphi^4$-model in the framework of the $\epsilon$-expansion in the 5-th order of perturbation theory and for the first time to carry out an independent verification of the results presented in~\cite{Chetyrkin1981,Chetyrkin1983,Kazakov1983, Kleinert1991}. It should be noted that the main drawback of sector decomposition technique is that the integration domain is divided into a huge number of sectors one needs to calculate to get the value of a particular diagram. The authors of~\cite{Adzhemyan2013five-loop} succeeded to reduce the number of sectors dramatically due to the most complete account of symmetries of diagrams.

While performing calculations in the framework of the renormalization group in the fixed space dimension, one needs to calculate integrals that contain no divergences. In this paper we use sector decomposition technique to calculate
the Fischer  exponent  $\eta$ in the six loop approximation in the space of fixed dimension $d = 2$ for the $\varphi^4$-model. We also describe all the technical details of the proposed method of calculation. Such a calculation is of current interest for two reasons. On one hand, it is well known that finding of the critical exponents in the model $\varphi^4$ in $d = 2$ is very difficult both in $\epsilon$-expansion, and in the fixed dimension space. On the other hand, there are the Onsager's exact solution  for the Ising model to compare with.

There are several results achieved in $\varphi^4$-model in the framework of fixed space dimension: for $d=3$ the seventh order of loop expansion was calculated~\cite{Nickel1977} (1977), for $d=2$ the fourth order of loop expansion was presented in the paper~\cite{Nickel1977} as well, and the latest available result is the fifth order of loop expansion that was published in~\cite{Orlov2000,Calabrese} (2000). The authors of the latter paper had also provided a resummation procedure and found that in the case of Ising model the results were still very far from the well known exact Onsager solution. This disagreement as well as the rapid development of numerical methods of diagram calculation were motivating reasons for us to perform the six loops calculations that this paper is devoted to.

The structure of the article is as follows. In section~\ref{sec_renorm}, we describe the general renormalization scheme we use. In section~\ref{sec_feynman} we recall the Feynman representation of the Green functions that our calculations are based upon. Section~\ref{sec_sd} is both essential and the most complicated part of the paper, it contains detailed description of our own implementation of the sector decomposition strategy, symmetry finding and $R'$-operation. In the last section~\ref{sec_results} we present the new results of the Fisher exponent $\eta$  calculation made with described technique as well as its Borel resummation. Discussion and conclusion are given in section~\ref{sec_conclusion}. In the Appendix~\ref{appendix_sd} we explain how to perform  the sector decomposition combined with $R'$-operation, in the Appendix~\ref{appendix_gstar} we explain why the coupling constant in the fixed point has the universal value in the renormalization scheme that is offered in this paper, and in Appendix~\ref{appendix_diags} there are the numerical results of the calculation presented for all the diagrams.

\section{Renormalization scheme}
\label{sec_renorm}
\input{renorm}

\section{Feynman representation}
\label{sec_feynman}
\input{feynman}

\section{Sector Decomposition and Graph Symmetries}
\label{sec_sd}
\input{sd}

\section{Results and resummation}
\label{sec_results}
\input{resummation}

\section{Conclusions/Discussion}
\label{sec_conclusion}
In this paper we presented the approach for computation of the critical exponents in models with fixed space dimension based on Bogoliubov-Parasiuk $R$-operation and sector decomposition technique. This approach allows one to easily automate such a computations and to perform generalization for more complicated types of symmetries. As an illustration of this approach the critical exponent $\eta$ (Fisher exponent) of the $O(N)$-symmetric $\varphi^4$ model in the fixed space dimension $d=2$ was calculated numerically up to  6 loop approximation.

In section \ref{sec_sd} we presented a modification of the sector decomposition strategy based on the Speer sectors which allows to utilize graph symmetries, thus reducing number of sectors we need to calculate. Such an improvement is very important for 6 and more loop diagrams, where the number of sectors for a particular diagram exceeds  $10^5$. Of course, comparing with  semi-analytical approach used in \cite{Nickel1977}, sector decomposition may be considered as a brute force method, but the main benefit here is that the only limiting factor here is the available computational power, so no other modifications are needed to compute 6 loop approximation of four-point function and so on. 
It also should be noted, that due to some specifics of the model, analytical tricks used in \cite{Nickel1977} are more developed for 3D case where 6 loop approximation was calculated in 1977 (contrary to 4 loops in 2D case). the number of such a tricks  in 2D case is very limited, but they are effective in 6 loops as well, so to get higher precision of the final result and minimize computational time we combine both semi-analytical methods from \cite{Nickel1977} and pure numerical methods discussed earlier (see discussion in Appendix~\ref{appendix_diags}).

\begin{figure}[h]
\includegraphics[width=14cm]{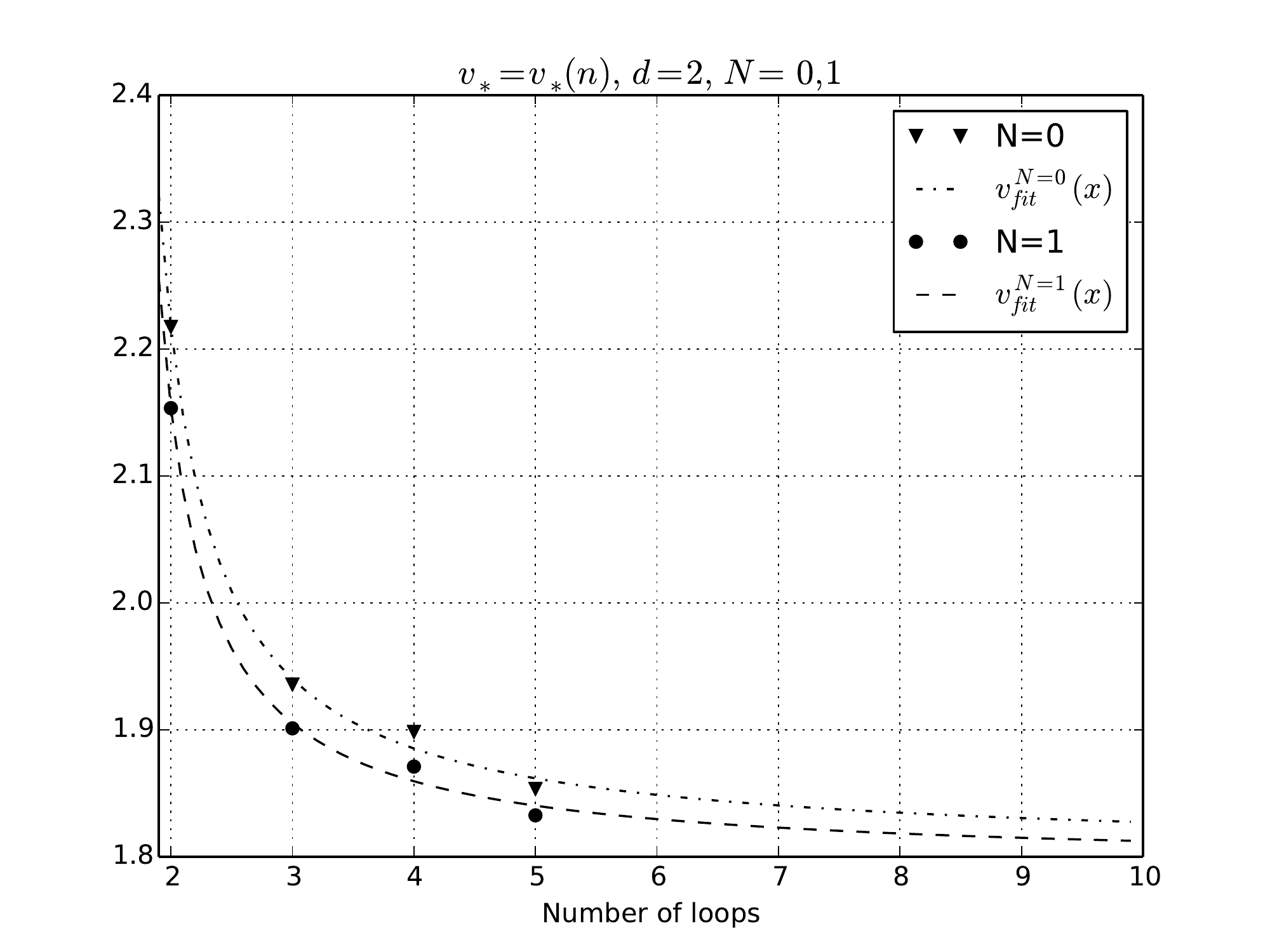}
\caption{Values of $v_*$ ($N=0,1$) for different number of loops and fit of this data with $v_{fit}=a/(x+b)+c$ (for $N=0$: $a=0.205$, $b=-1.51$, $c=1.80$, for $N=1$ $a=0.162$, $b=-1.55$, $c=1.79$).}
\label{fig:vstar}
\end{figure}

\begin{figure}[h]
\includegraphics[width=14cm]{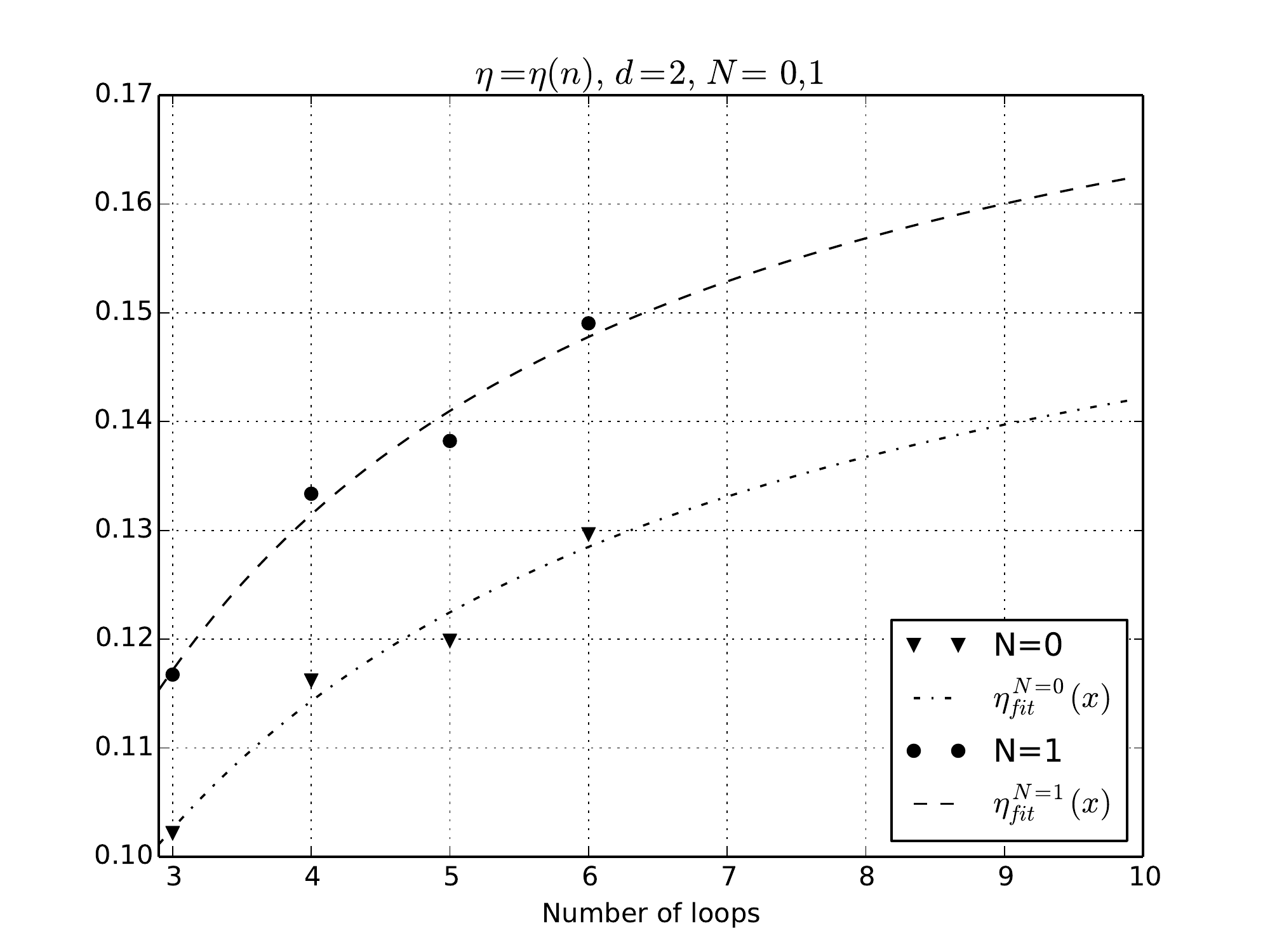}
\caption{Values of the Fisher exponent $\eta$ ($N=0,1$) for different number of loops and fit of this data with $\eta_{fit}=a/(x+b)+c$ (for $N=0$: $a=-0.240$, $b=1.04$, $c=0.137$, for $N=1$ $a=-0.342$, $b=1.17$, $c=0.181$).}
\label{fig:eta}
\end{figure}

Resummed values for $v_*$ and $\eta$ presented in Tables \ref{tab:vstar} and \ref{tab:eta} show  correct tendencies (with higher number of loops taken into account resummed values are closer to that predicted by high temperature expansion and exact solution correspondingly). In the figure~\ref{fig:vstar} the value of the fixed point is drawn with respect to number of loops taken into account for $N=0$ and $N=1$, the dashed lines are fit of this values with function $v_{fit}=a/(x+b)+c$  (for $N=0$: $a=0.205$, $b=-1.51$, $c=1.80$, for $N=1$ $a=0.162$, $b=-1.55$, $c=1.79$).The analogous plot for $\eta$ is presented on the figure~\ref{fig:eta}, dashed lines correspond to the fit with  $\eta_{fit}=a/(x+b)+c$ (for $N=0$: $a=-0.240$, $b=1.04$, $c=0.137$, for $N=1$ $a=-0.342$, $b=1.17$, $c=0.181$). 

Of course, asymptotic values of the fits on the figures \ref{fig:vstar}, \ref{fig:eta}  cannot be considered as some estimation for the real values of $v_*$ and $\eta$ because we don't have enough points for accurate fit and also the particular choice of the fitting function is almost arbitrary. But these fits show us the tendency for the higher order contribution which goes in the right direction: 6-loop contribution has led to some increase in the value of Fischer exponent ($\eta_6 = 0.1490$ vs $\eta_5 = 0.1382$ in 5-loop approximation), which is in agreement with the dynamics of the exponent growth in the previous approximations. However, this value is still significantly smaller than $\eta = 0.25$ in the exactly solvable 2D Ising model. Such a discrepancy is usually interpreted as an effect produced by non-analytical terms \cite{Orlov2000}.

\begin{acknowledgements}
 Authors would like to thank \textit{ N.V.~Antonov}, \textit{A.I.~Sokolov} and \textit{E. Zerner-K{\"a}ning} for fruitful discussions. We acknowledge Saint-Petersburg State University for a research grant 11.38.185.2014.
We also thank Resource Center ”Computer Center of SPbU” and Far-Eastern Federal University for providing computational resources.
\end{acknowledgements}

\appendix
\section{Sector decomposition: neat example}
\label{appendix_sd}
\input{appendix_sd}

\section{Universal amplitude ratio and the charge value in the fixed point in $d=2$}
\label{appendix_gstar}
\input{appendix_gstar}

\section{Results for individual diagrams}
\label{appendix_diags}

The contents of this Appendix provide all the necessary information to calculate renormalization constants of the $O(n)$ symmetric $\varphi^4$ theory.  
An individual diagram $\gamma$ is defined by the Nickel index (see~\cite{Nickel1977}).  The contribution of $\gamma$ into normalized 1-PI Green functions $\bar{\Gamma}_i$ (see~\eqref{bar_Gamma}) is denoted as $\bar{\gamma}$. For $\bar{\Gamma}_4$ the value of $\bar{\gamma}$ coincides with $\gamma$. For $\bar{\Gamma}_2$ the value of $\bar{\gamma}$ is defined by the derivative of $\gamma$ with respect to the square of external momentum: $\bar{\gamma} = -\partial_{p^2}\gamma$. The following tables contain numerical results of calculation in the same normalization as in~\cite{Nickel1977}.

The structure of tables is as follows: the number of diagram $\gamma$ is in first column, the second column gives Nickel index of the corresponding diagram.  In the third column we give calculated results of the contribution of $\gamma$ into $\bar{\Gamma}_i$ with the account of subtractions of self energy subgraphs (see~\eqref{R_2},\eqref{k2} in the Section~\ref{sec_renorm}; note that values in this table differ from those presented in \cite{Nickel1977} due to the different self energy subgraphs subtraction procedure, see discussion after \eqref{diagsum}). The value of $KR'$ of the graphs are presented in the fourth column. The fifth column contains symmetry coefficients, and the last column contains additional combinatorial factors $O_N(\gamma)$ for $O(N)$-symmetric $\varphi^4$ model. 

Crosses in the first column of tables denote the factorizable diagrams (i.e. diagrams which are simple product of diagrams with lower loop count). There is no need to calculate such diagrams using sector decomposition, because lower loop graphs are calculated with higher precision. The asterisks in the first column mark diagrams which can be easily calculated using a number of analytical tricks which allow one to reduce these diagrams to integrals with lower loop count. The most simple trick is to integrate out all one loop bubble subgraphs using analytical expression for massive bubble on external momenta $\mathbf{k}$ ($x\equiv \mathbf{k}^2$):
\begin{equation}
F(x)=\frac{\log[1+2x+x^2/2+(1+x/2)(x(4+x))^{1/2}]}{(x(4+x))^{1/2}}, \quad F(0)=1.
\end{equation}
Latter allows to calculate a set of diagrams with precision unavailable for straightforward calculation.

New results – the values of 6 loop diagrams for $\bar{\Gamma}_2$ – are given in Table~\ref{table_V}. Results of the calculation of all diagrams with the number of loops $n\le5$ are given in Tables~\ref{table_III} and \ref{table_IV}. The results for the diagrams with the account of the subtraction of self energy diagrams (the third column) obtained by us for $n\le5$ are the same (within the given error) as in~\cite{Orlov2000,Calabrese, Sokolov_privatecomm}, and for $n\le4$ all diagrams without self energy subgraphs are the same as in~\cite{Nickel1977}.

Diagram values in the fourth column (result of $R'$ operation) in the presented form might be useful in the sense that one can easily generalize them for systems with more complicated structure, in particular,  using additional combinatorial factors $O_N(\gamma)$ (the last column).

\subsection{Results up to 5 loops}
Results of the calculation of all vertex diagrams with the number of loops $n=1,\dots,5$ are given in Table~\ref{table_III}. 
Table~\ref{table_IV} contains the results of all self energy diagrams with up to 5 loops.

\input{diagTable_5loops}

\subsection{Results, 6 loops}
Table~\ref{table_V} gives the results of the calculation of all self energy diagrams in six loops. 
\input{diagTable}

\bibliographystyle{unsrt}
\bibliography{main}

\end{document}

%% file: renorm.tex
In this section we discuss the renormalization scheme that we used. This scheme is formulated in a different way than one used in~\cite{Nickel1978, Orlov2000}, but actually produces the same renormalization constants. The reason for this is that it is more convenient for us to keep renormalization mass in the action and use the renormalization group equations rather than the Callan-Symanzik equations.

 The questions discussed in this section are quite general, and everything is valid for both 2D and 3D $\varphi^4$ model, so we will keep $d$ as space dimension in all equations in this section, keeping in mind that $d=2$ or $d=3$.
 
Let us consider the scalar $\varphi^4$ model ($N=1$) with the following renormalized action ($O(N)$-symmetric generalization of the model discussed at the end of this section):
\begin{equation}
\label{SR}
S=-\frac{1}{2}(m^2 Z_1+k^2 Z_2+\delta m^2)\varphi^2-\frac{1}{24}g
\mu^{{(4-d)}}Z_3\varphi^4,
\end{equation}
where the constants $Z_i$ are related to the renormalization constants of the mass $Z_{m^2}$, the field $Z_\varphi$, and the charge $Z_g$:
\begin{equation}
\label{ZZ}
 Z_1=Z_{m^2}Z_\varphi^2, \quad Z_2=Z_\varphi^2, \quad Z_3=Z_g Z_\varphi^4.	
\end{equation}
Here $Z_{m^2}$, $Z_{\varphi}$ and $Z_g$ are defined as:
\be
\label{masses}
m^2_0=m^2\,Z_{m^2},\quad g_0=g\mu^{4-d}Z_g,
\quad\varphi_0=\varphi Z_\varphi.
\ee
The variables with subscript $0$ are bare parameters, $\mu$ is the renormalization mass.

In~\cite{Nickel1978, Orlov2000} subtraction at zero momenta (ZM) was used, 
the latter one has no renormalization mass $\mu$ in the action~\eqref{SR} ($\mu$ is replaced by  $m$) and instead of RG-equations~\eqref{RGeq} Callan-Symanzik equations are used. Moreover $m^2$ in this scheme is not proportional to $|T-T_c|$. In this paper we will use the renormalization scheme which is more convenient for us. On one hand this scheme produces exactly the same renormalization constants as one used in~\cite{Nickel1978, Orlov2000}, on the other hand it allows us to use standard renormalization group equations. 

Renormalization scheme used in this paper is defined by the following normalization conditions

\begin{equation} \label{cond}
\begin{aligned}
&{\Gamma_2^R|_{p=0, \mu=m} =-m^2,} & \Gamma_2^R|_{p=0, m=0} &= 0, \\
\partial_{p^2}&\Gamma_2^R|_{p=0, \mu=m}=-1, & \Gamma_4^R|_{p=0, \mu=m} &= -gm^{(4-d)},
\end{aligned}
\end{equation}
where $\Gamma_2$, $\Gamma_4$ are the two-legged (self energy diagrams) and the four-legged (vertex type diagrams) one-particle irreducible (1-PI) 
Green functions, and $ \Gamma_2^R$ and $\Gamma_4^R$ are their renormalized analogues. 
In terms of normalized functions 
\begin{equation}
\label{bar_Gamma}
\bar \Gamma_2=-\partial_{p^2}\Gamma_2 , \quad \bar
\Gamma_4=-\frac{\Gamma_4}{g\mu^{(4-d)}}
\end{equation}
the conditions \eqref{cond} take the form
${\bar \Gamma_i^R|_{p=0, \mu=m}=1}$, ${i=2,4}$. As in minimum subtractions (MS) scheme, in renormalization scheme that we use in this paper all renormalization constants depend only on dimensionless charge $g$ (and are independent of $m$).

Renormalization group (RG) equations are derived in a standard way 
(on the basis of arbitrariness of renormalization mass $\mu$) in our renormalization scheme.
They have the same form as in the MS scheme:
\begin{equation}\label{RGeq}
 (\mu\partial_\mu+\beta \partial_g-\gamma_{m^2} m^2
\partial_{m^2})\Gamma_n^R=n\gamma_\varphi \Gamma_n^R\,, 
\end{equation}
where
\begin{equation}
\label{27}
 \beta=\frac{-{(4-d)}\,g}{1+g\partial_g \ln Z_g}\,, \qquad \gamma_i=\frac{-{(4-d)}\,g\partial_g \ln Z_i}{1+g\partial_g \ln Z_g}\,.
\end{equation}
At the fixed point $\beta(\gStar)=0$, RG-equations turn into equations of critical 
scaling with anomalous exponents $\gamma_i(\gStar)$. In particular,  $\gamma_2(\gStar)=\eta$ is the critical Fisher exponent.

Renormalization constants are constructed using $R'$ operation, 
which is incomplete Bogoliubov-Parasiuk $R$-operation 
that carries out subtractions in subgraphs.
The renormalization constants look in this notation as 
\be \label{ZR}
Z_2= 1 - K^{(0)}R'\bar\Gamma_2,\qquad
Z_3=1-K^{(0)}R'\bar \Gamma_4,
\ee
where $K^{(0)}$ operation in~\eqref{ZR} in our renormalization scheme looks like:
\be
 K^{(0)}\Gamma=\Gamma|_{p=0, \mu=m},
  \label{k0}
\ee
i.e., we subtract the initial part of the Taylor expansion in the external momenta, and additionally we assume $\mu=m$.

Calculation of the $R'$-operation of a diagram can be split into two stages: $R'\Gamma_i=R_4'R_2'\Gamma_i$. 
First we perform subtraction of the self energy subgraphs using $R_2'$:
\be
\label{R_2}
\Gamma _{i}^{(2)} \equiv R_{2}^{'} \, \Gamma _{i} =\prod \limits _{j}\left(1-K_{j}^{(2)} \right) \, \Gamma _{i}, 
\ee
where the product is taken over all the self energy subgraphs, and $K_{j}^{(2)}$-operation allocates from $j$-th subgraph two terms of its Taylor expansion in the square of external momenta $q_{j}^{2} $:
\be
K^{(2)}_j\Gamma=\Gamma|_{q_j=0}+q_j^2\left(\partial_{q_j^2}\Gamma|_{q_j=0}\right).
 \label{k2}
\ee
Such subtractions can be expressed as a remainder of the Taylor expansion
\be
\label{1-K}
\left(1-K^{(2)} \right)\, f(q^{2} )= \int \limits _{0}^{1}da\, (1-a)\,  \partial _{a}^{2} f(aq^{2}).
\ee
At the second step, using $R_4'$ which finalizes $R'$-operation, we subtract vertex subgraphs from $\Gamma_{i}^{(2)}$ (see~\cite{Vasilev2004} for details):
\be
\label{L-operation}
R'\, \Gamma _{i} =R_{4}^{'} \Gamma _{i}^{(2)} =\left(1{-}\sum \limits _{j}L_{j}  {+}\sum \limits _{j,l}L_{j} L_{l} {-}\dots\right)\Gamma _{i}^{(2)},\quad
L_j\equiv(K^{(0)}R')_j\,,
\ee
where a single summation is taken over all the different vertex subgraph $j$, double summation -- over all pairs of disjoint vertex subgraphs $j,l$, then triple summation etc. 
For examples of $R'_4$ operation see section~\ref{R-prime}.

To calculate renormalization constants~\eqref{ZR} for scalar $N=1$ model we need to calculate $K^{(0)}R'\bar{\Gamma}_i$, which can be represented by the following perturbation series:
\be
K^{(0)}R'\bar{\Gamma}_i =\sum\limits_n (-u)^n\,K^{(0)}R'{G}_i^{(n)} 
=\sum\limits_n (-u)^n
	\sum\limits_{\bar{\gamma}\in G_i^{(n)}} Sym(\bar{\gamma})\times  K^{(0)}R'(\bar{\gamma}),
\label{diagsumn1}
\ee
where ${G}_i^{(n)}$ -- $n$-loop term of $\bar{\Gamma}_i$, $\bar{\gamma}$ enumerates diagrams in ${G}_i^{(n)}$, $Sym(\bar{\gamma})$ is the symmetry factor of a particular diagram in ${G}_i^{(n)}$, $K^{(0)}R'(\bar{\gamma})$ -- counterterm to $\bar{\gamma}$.    In our calculation for $d=2$ we use $u = g/(4\pi)$, according to~\cite{Nickel1977} we normalize integrals by factor $1/\pi$ for each loop integration.

In the case of $O(N)$-symmetric model the field has $N$ components:  $\varphi_a(x), a=1..N$, and  one should read $\varphi^2$ as  $\sum_{a=1}^N \varphi_a\varphi_a$ and $\varphi^4$ as $(\varphi^2)^2$. Latter can be written in symmetrized form as follows  $(\varphi^2)^2=\sum_{a,b,c,d=1}^N V_{abcd}  \,\varphi_a\varphi_b\varphi_c\varphi_d$ with vertex factor $ V_{abcd}=(\delta_{ab}\delta_{cd}+\delta_{ac}\delta_{bd}+\delta_{ad}\delta_{bc})/3$. Each diagram line now has additional factor $\delta_{ab}$ and each vertex has factor $V_{abcd}$. All these additional factors can be factorized out from the integral and as a result for the $O(N)$-symmetric case we got an additional $N$-component combinatorial factor $O_N(\bar\gamma)$:
\be
K^{(0)}R'\bar{\Gamma}_i =\sum\limits_n (-u)^n\,K^{(0)}R'{G}_i^{(n)} 
=\sum\limits_n (-u)^n
	\sum\limits_{\bar{\gamma}\in G_i^{(n)}} Sym(\bar{\gamma})\times  K^{(0)}R'(\bar{\gamma})\times O_N(\bar{\gamma}),
\label{diagsum}
\ee
here $Sym(\bar{\gamma})$ and $K^{(0)}R'(\bar{\gamma})$ are the same as for scalar~($N=1$) $\phi^4$ model.
Numerical values that were calculated for $Sym(\gamma)$, $KR'(\gamma)$ and  $O_N(\gamma)$ are listed in Appendix~\ref{appendix_diags}. 

In the same manner one can easily generalize scalar $\varphi^4$ model \eqref{SR} and eq.\eqref{diagsumn1} for more complicated types of symmetries~\cite{Antonov2013, NalimovT3}. 

It should be noted, that authors ~\cite{Nickel1977, Nickel1978, Orlov2000} also calculated diagram values using the self energy subgraphs subtractions, but  authors ~\cite{Nickel1977, Nickel1978}  subtract from the diagrams only self energy subgraphs at zero external momenta (without the term proportional to the square of external momenta), while authors \cite{Orlov2000} use the same subgraph subtraction procedure as in this paper, and thus calculate values for $\Gamma_i^{(2)}$. Moreover, the authors ~\cite{Nickel1977, Nickel1978, Orlov2000} didn't use representations~\eqref{R_2} and~\eqref{1-K} for subtraction procedure, which are convenient for our computer implementation, as well as $R'$ operation for representation of the renormalization constants~\eqref{ZR},\eqref{diagsum}.
And, as it was mentioned above,  this approach allows one to  easily generalize scalar $O(1)$ model to different types of symmetries.

%% file: feynman.tex
Calculations were performed in the Feynman representation using sector decomposition technique~\cite{Heinrich2000}.
This technique was originally used for the analysis of ultraviolet (UV) divergences in the proof of the Bogoliubov-Parasiuk theorem on renormalization~\cite{Hepp1966}. 
But despite the fact that there are no divergences in  diagrams of considered theory ($d=2$), using the sector decomposition technique is
justified by the advantages noted above: the resulting integrals converge well 
and can be easily calculated numerically.

We recall some useful facts about the Feynman representation. Diagrams of the
Green functions $\Gamma_2$ and $\Gamma_4$ in the $n$-loop approximation correspond to integrals of the form:
\be
\label{feyn_int}
J_{n} =\int \frac{d\mathbf{ k}_{1}}{\left(2\pi \right)^{d}}  \dots\int \frac{d\mathbf{ k}_{n}}{\left(2\pi \right)^{d}}
    \prod \limits _{i=1}^{s} \frac{1}{\left({\mathbf{ q_i}}^{2} +m^{2} \right)^{\lambda _{i} }},
\ee
where $\lambda _{i}$ -- the number of edges in the diagram with a momentum $\mathbf{ q}_{i}$, $s$ -- the number of edges with different momenta $\mathbf{ q}_{i}$, where $\mathbf{ q}_{i}$ are linear combinations of integration momenta $\mathbf{ k}_{j}$ and, for the self energy diagrams, external momentum $\mathbf{ p}$. Vertex diagrams will be considered at zero external momenta.
The total number of edges $\alpha =\sum \limits _{i}\lambda _{i}$ in the self energy diagrams is equal to $\alpha =2n-1$,  in the vertex diagrams -- to $\alpha =2n$.
Sum of momenta $\mathbf{ q}_{i}$ flowing into each vertex is equal to zero («the law of momentum conservation»). Using the Feynman formula, we obtain:
\be
\label{feyn_repr}
\prod \limits _{i=1}^{s}\frac{1}{\left(\mathbf{ q}_i^{2} +m^{2} \right)^{\lambda _{i} } }  =
    \frac{\Gamma \left(\alpha \right)}{\prod \limits _{i=1}^{s}\Gamma \left(\lambda _{i} \right) } 
        \int \limits _{0}^{1} \dots \int \limits _{0}^{1}du_{1} \dots du_{s}   \frac{\delta \left(\sum \limits _{i=1}^{s}u_{i}  -1\right)
        \prod \limits_{i=1}^{s}u_{i}^{\lambda _{i} -1}}  {\left[\sum \limits _{i=1}^{s}\left(\mathbf{ q}_i^{2} +m^{2} \right)u_{i}  \right]^{\alpha } },
\ee
where the number $s$ of \textit{Feynman parameters} $u_i$ is equal to the number of different momenta in~\eqref{feyn_int}.

The denominator of the integrand in \eqref{feyn_repr} is quadratic form of momenta $\mathbf{ k}_{j}$ in power $\alpha$:
\be
\sum\limits_{i=1}^{s}\left(\mathbf{ q_i}^{2} +m^{2} \right)u_{i}  =\sum\limits_{i=1}^{s}\sum\limits_{j=1}^{s}v_{ij} \mathbf{ k}_{i} \mathbf{ k}_{j} +2\sum\limits_{i=1}^{s}\mathbf{ b}_{i} \mathbf{ k}_{i} +c,
\ee
where vectors $\mathbf{ b}_{i}$ are determined by external momenta of the diagram. The integral of the quadratic form  in some power is known:
\be
\label{feyn_quad_form}
\left(2\pi \right)^{-dn} \int \dots \int \frac{d\mathbf{ k}_{1} \dots d\mathbf{ k}_{n}}{\left[v_{ij} \mathbf{ k}_{i} \mathbf{ k}_{j} +2\mathbf{ b}_{i} \mathbf{ k}_{i} +c\right]^{\alpha }}=
\frac{\left(4\pi \right)^{-dn/2} \Gamma \left(\alpha -dn/2\right)}{\Gamma \left(\alpha \right)\left[c-\left(v^{-1} \right)_{ij} \mathbf{ b}_{i} \mathbf{ b}_{j} \right]^{\alpha -dn/2} \left(\det v\right)^{d/2} }.
\ee
Considering \eqref{feyn_repr}--\eqref{feyn_quad_form}, we obtain from \eqref{feyn_int} 
\be
\label{feyn_int_feyn_repr}
J_{n} = \frac{\left(4\pi \right)^{-dn/2} \Gamma \left(\alpha -dn/2\right)}{\prod \limits _{i=1}^{s}\Gamma \left(\lambda_{i} \right)} 
    \int \limits _{0}^{1}\dots \int \limits _{0}^{1}du_{1} \dots du_{s}
        \frac{\delta \left(\sum \limits _{i=1}^{s}u_{i} -1 \right)
            \prod \limits _{i=1}^{s}u_{i}^{\lambda _{i} -1}}  
                {\left[c-\left(v^{-1} \right)_{ij} \mathbf{ b}_{i} \mathbf{b}_{j} \right]^{\alpha -dn/2} \left(\det v\right)^{d/2}}.
\ee
For vertex diagrams at zero external momentum $\mathbf{ b}_{i} =0$, $c=m^{2}$  
(taking into account $\sum \limits _{i=1}^{s}u_{i} =1$), $\alpha =2n$, then
\be
\label{J_4}
J_{n}^{(4)} =\frac{\left(4\pi \right)^{-dn/2} \Gamma \left(n(2-d/2)\right)m^{n(d-4)}} 
    {\prod \limits _{i=1}^{s}\Gamma \left(\lambda _{i} \right) } 
    \int \limits _{0}^{1}\dots \int \limits _{0}^{1}du_{1} \dots du_{s}
    \frac{\delta \left(\sum \limits _{i=1}^{s}u_{i} -1\right)
        \prod \limits _{i=1}^{s}u_{i}^{\lambda _{i} -1}}  {\left(\det v\right)^{d/2}}.
\ee 
For self energy diagrams $\alpha =2n-1$, the quantities $\mathbf{ b}_{i}$ and $c$ take the form $\mathbf{ b}_{i} =c_{i} \mathbf{ p}_{i}$, $c=m^{2} +hp^{2}$, then   from \eqref{feyn_int_feyn_repr} we obtain
\be
\label{J_2}
\left.\partial_{p^{2} } J_{n}^{(2)}\right|_{p=0} =
    -\frac{\left(4\pi \right)^{-dn/2} \Gamma \left(n(2-\frac{d}{2})\right)m^{n(d-4)}} {\prod \limits _{i=1}^{s}\Gamma \left(\lambda _{i} \right)} 
    \int \limits _{0}^{1}\dots \int \limits _{0}^{1}du_{1} \dots du_{s} 
        \frac{\delta \left(\sum \limits _{i=1}^{s}u_{i}  -1\right)\prod \limits _{i=1}^{s}u_{i}^{\lambda _{i} -1}{\cdot}Q}{\left(\det v\right)^{d/2+1} },
\ee
where we introduced a new notation 
\be
\label{feyn_Q}
Q\equiv \left[h-\left(v^{-1} \right)_{ij} \mathbf{ c}_{i} \mathbf{ c}_{j} \right]\det v.
\ee

Dependence of $\det v$ and $Q$ on Feynman parameters $u_{i}$ is not related to the choice of momentum flow and can be found directly from the appearance of the diagram itself, bypassing the above procedure of transition from the momentum representation (see, e.g.,~\cite{zavialov}).
We used the following rules to determine both $\det v$ and $Q$:
\begin{itemize}
\label{Rules}
\item Determinant $\det v$ of the $n$-loop diagram is equal to the sum of products of Feynman parameters $u_{i}$, each of which contains $n$ distinct factors.
\item Terms, that contain a subset of parameters for which the sum of the corresponding momenta is zero (conservation law), are prohibited.
\end{itemize}

\begin{figure}

\begin{tikzpicture}[node distance=2cm]
\coordinate[vertex] (v1); 
\coordinate[vertex, right=of v1] (v2);
\coordinate[vertex, right=of v2] (v3);
\coordinate[vertex, right=of v3] (v4);
\coordinate(x1) at ($ (v2)!.5!(v3) $);
\coordinate[vertex] (v5) at ($ (x1)!1.4!270:(v3) $);
 
\coordinate[right=of v4] (xe2);
\coordinate[left=of v1] (xe1);
\coordinate[below left=of v5] (xe3);
\coordinate[below right=of v5] (xe4);
\coordinate(e2) at ($ (v4)!.15!(xe2) $);
\coordinate(e1) at ($ (v1)!.15!(xe1) $);
\coordinate(e3) at ($ (v5)!.15!(xe3) $);
\coordinate(e4) at ($ (v5)!.15!(xe4) $);

\arcll{v2}{v1}{45}{3};
\arcrl{v2}{v1}{45}{2};

\arcll{v3}{v2}{45}{5};
\arcrl{v3}{v2}{45}{4};

\arcll{v4}{v3}{45}{7};
\arcrl{v4}{v3}{45}{6};

\draw (v1) -- (e1);
\draw (v4) -- (e2);
\draw (v5) -- (e3);
\draw (v5) -- (e4);
\arcll{v5}{v1}{25}{1};
\arcrl{v5}{v4}{25}{1};

\end{tikzpicture}
\caption{Sample vertex diagram ($ee12|e33|e44|44||$). Number on edge denotes the Feynman parameter corresponding to this edge. Note that if external momenta is set to zero both lines marked by \textbf{1} have the same momenta and thus the same Feynman parameter with $\lambda_1 =2$}
\label{4loop4leg}
\end{figure}



Let us  consider this rule for the diagram  at zero external momenta plotted on Fig.\ref{4loop4leg}. It is clear that here $\lambda _{1} =2$ since there are two edges that correspond to the parameter $u_1$, for other Feynman parameters $\lambda _{2} =\lambda _{3} =\lambda_{4} =\lambda _{5} =\lambda _{6} =\lambda _{7} =1$.
The determinant represents a sum of monomials which are products of the four (the number of loops) different Feynman parameters. There are $\binom{7}{4}=35$ possible different monomials here. We should exclude from them the monomials which are prohibited by the conservation laws.
The straightforward conservation laws in vertices prohibit monomials which contain product $u_{1} u_{2} u_{3}$,
monomials $u_{2} u_{3} u_{4} u_{5}$ and $u_{4} u_{5} u_{6} u_{7}$, 
and also monomials which contain product $u_{1} u_{6} u_{7}$.
Additionally, it is necessary to take into consideration the composite conservation laws which prohibit monomial  $u_{2}u_{3}u_{6}u_{7}$ and  monomials with $u_{1} u_{4}u_{5}$. 
Thus, we have 20 products in the remainder, which define the determinant:
\be
\label{feyn_det_1}
\begin{aligned}
\det v=& u_{1} u_{2} u_{4} u_{5} +u_{1} u_{2} u_{4} u_{6} +u_{1} u_{2} u_{4} u_{7} 
        +u_{1} u_{2} u_{5} u_{6} +u_{1} u_{2} u_{5} u_{7} +u_{1} u_{3} u_{4} u_{6} \\
    +&u_{1} u_{3} u_{4} u_{7} +u_{1} u_{3} u_{5} u_{6} +u_{1} u_{3} u_{5} u_{7} 
        +u_{2} u_{3} u_{4} u_{6} +u_{2} u_{3} u_{4} u_{7} +u_{2} u_{3} u_{5} u_{6} \\
    +&u_{2} u_{3} u_{5} u_{7} + u_{2} u_{4} u_{5} u_{6} +u_{2} u_{4} u_{5} u_{7} +u_{2} u_{5} u_{6} u_{7} +u_{3} u_{4} u_{5} u_{6} +u_{3} u_{4} u_{5} u_{7} \\
    +&u_{3} u_{4} u_{6} u_{7} +u_{3} u_{5} u_{6} u_{7}.
\end{aligned}
\ee

The rule for constructing the determinant for the self energy (two-tailed diagrams)  
is the same as for the four-tailed ones.
For them it is necessary to know the value of $Q$ from~\eqref{feyn_Q} as well. It is built according to the rule, similar to the rule for constructing the determinant.
The difference is that the number of co-factors $u_{i}$ is incremented by one (i.e., equal to $n +1$ for $n$-loop diagram)
and we do not need to take into account the conservation laws in the external vertices and all composite conservation laws which include one of the external vertices.

Consider this rule for  the diagram drawn on Fig.\ref{3loop2leg}.
\begin{figure}[h]
\begin{tikzpicture}[node distance=2cm]
\coordinate[vertex] (v1); 
\coordinate[vertex, right=of v1] (v2);
\coordinate[vertex, right=of v2] (v3);
 
\coordinate[right=of v3] (xe2);
\coordinate[left=of v1] (xe1);
\coordinate(e2) at ($ (v3)!.15!(xe2) $);
\coordinate(e1) at ($ (v1)!.15!(xe1) $);

\arcrl{v1}{v3}{55}{1};

\arcll{v2}{v1}{35}{3};
\arcrl{v2}{v1}{35}{2};

\arcll{v3}{v2}{35}{5};
\arcrl{v3}{v2}{35}{4};

\draw (v1) -- (e1);
\draw (v3) -- (e2);
\end{tikzpicture}
\caption{Sample self-energy diagram ($e112|22|e|$). Number on edge denotes the Feynman parameter corresponding to this edge. }
\label{3loop2leg}
\end{figure}
The only law of conservation prohibits the presence of a product $u_{2} u_{3} u_{4} u_{5}$ in $Q$, thus we have:
$$Q=u_{1} u_{2} u_{3} u_{4} +u_{1} u_{2} u_{3} u_{5} +u_{1} u_{2} u_{4} u_{5} +u_{1} u_{3} u_{4} u_{5}. $$ 
While constructing the determinant we should also consider the conservation laws in the external vertices, that prohibit products $u_{1} u_{2} u_{3} $ and $u_{1} u_{4} u_{5}$, and so we obtain
\be
\label{feyn_det_2}
\begin{aligned}
\det v=&u_{1} u_{2} u_{4} + \cancel{u_1 u_2 u_3} + u_{1} u_{2} u_{5} +u_{1} u_{4} u_{3} + \cancel{u_1 u_4 u_5} + u_{1} u_{3} u_{5} +u_{2} u_{4} u_{3} +u_{2} u_{4} u_{5} +u_{2} u_{3} u_{5} +u_{4} u_{3} u_{5} = \\
=&u_{1} u_{2} u_{4} +u_{1} u_{2} u_{5} +u_{1} u_{4} u_{3} +u_{1} u_{3} u_{5} +u_{2} u_{4} u_{3} +u_{2} u_{4} u_{5} +u_{2} u_{3} u_{5} +u_{4} u_{3} u_{5}.
\end{aligned}
\ee
The alternative way of constructing Feynman representation from graph is to use 1- and 2-trees~\cite{zavialov}; approach presented here is fully equivalent to it, but from our point of view it is more intuitive and allows simpler implementation.

%% file: sd.tex
\subsection{Sector decomposition method}
\label{sd_scheme}

Originally sector decomposition method was designed to extract divergences from the integrals corresponding to the Feynman graphs. In this paper we consider only finite integrals without divergences and will discuss only case of convergent integrals (for more details on extracting divergences see e.g.~\cite{Heinrich2008}).

When we try to calculate convergent graphs in Feynman representation numerically, we encounter a problem with the convergence of numerical integration of these integrals due to integrable singularities, which are arising as a consequence of zeroes of $\det v$ in~\eqref{J_4},~\eqref{J_2} over the sets of Feynman parameters.
The purpose of the sector decomposition is to split the integration domain over Feynman parameters $u_i$ into subdomains (\textit {sectors}) followed by  the substitution of variables in each sector in a way that zeros of $\det v$ are represented as powers of some certain variables and are canceled by the integration measure. The solution is based on the following partition of the integral over $k$-dimensional unit cube into the sum of $k$ integrals:

\be
\label{sd_1}
J = \int\limits_0^1dx_1...\int\limits_0^1dx_j...\int\limits_0^1dx_k\,f(x_1,...,x_j,...,x_k) = 
\sum\limits_{j=1}^k\int\limits_0^1dx_j\int\limits_0^{x_j}dx_1...\int\limits_0^{x_j}dx_{j-1}\int\limits_0^{x_j}dx_{j+1}...\int\limits_0^{x_j}dx_k\,f(x_1,...,x_j,...,x_k).
\ee
Replacing each variable in the right hand side as $x_l\to x_j\tilde{x}_l, l\neq j$ one can reduce the domain of integration to $[0,1]^n$. We omit tildes above the new variables of integration, and the result is:

\be
\label{sd_2}
J = \sum\limits_{j=1}^k\int\limits_0^1dx_1\dots\int\limits_0^1dx_{j}\dots\int\limits_0^1dx_k\,f(x_1 x_j\,,\dots,\,x_j\,,\dots,x_k x_j)\cdot x_j^{k-1}.
\ee

At each step of decomposition some set of Feynman parameters $\{u\}$ is chosen as a collection of variables $\{x\}$ in~\eqref{sd_1}. We will call this set a «decomposition domain», variable $u_j$ in $j$-th term of the sum will be called «main»; the rest of the variables from decomposition domain we will denote as «secondary» and the variables that do not belong to decomposition domain are «neutral». 
At the first step the full set of $s$ Feynman variables $\{u\}=\{u_1,\dots,u_s\}$ usually forms the decomposition domain. Determinant $\det v(u_1,\dots,u_s)$ consists of the sum of monomials over variables $u_i$ that share the same power $n$ (equal to number of loops). 
After substitution $u_l=u_{j_1}\tilde{u}_l,\ l\neq j_1$ and the consequent $\tilde{u}_l\to u_l$, it is possible to factor out the common factor $u_{j_1}^n$. Denoting the resulting sum as $\det_{j_1} v(u_1,\dots,u_s)$ \footnote{it really makes sense of determinant of the appropriate matrix}  we obtain:
\be
\label{sd_3}
\det v(u_1,\dots,u_s) \to \det v(u_1 u_{j_1},u_2 u_{j_1},...,  u_{j_1},..., u_s u_{j_1}) =  u_{j_1}^{n} \textstyle{\det_{j_1}} v(u_1,\dots,u_s).
\ee
As the result of the first step, the power of all monomials that contained factor $u_{j_1}$ in $\det v$ have been decreased by one and in $\det_{j_1} v$ became equal to $n-1$. The subsequent steps are intended for the analogous transformation of the determinant, which decrease minimal power of monomials it consists of.

The choice of decomposition domains that leads to the desired result is ambiguous. The algorithm of subsequent choice of decomposition domains is called \textit{strategy}.
There are a number of strategies developed  by different authors, see review part in~\cite{Carter2011}. We use our own implementation of Speer sectors~\cite{Speer1977}; the reason for choosing this strategy is that it produces a minimal set of final sectors for a given problem and allows to maximally take into account the symmetries of the graph for further reduction of the number of sectors to be calculated. The strategy described below is reasonable in relation to the problem of calculating diagrams at zero external momenta.

The strategy we use consists of $n$ steps, and each sector is defined by the set of variables $[j_1, j_2,\dots,j_n]$ which were «main» at successive steps of the procedure.
After $n$ steps it leads to the following transformation of determinant in the sector $[j_1,j_2,\dots,j_n]$:
\be
\label{sd_4}
\det v(u_1,\dots,u_s) \to u_{j_1}^{n}u_{j_2}^{n-1}\dots u_{j_n}\textstyle{\det_{j_1,j_2,\dots,j_n}} v(u_1,\dots,u_s).
\ee
The choice of decomposition domain at each step assumes that we should exclude «main» variables $u_{j_1},u_{j_2},\dots$  used in the previous steps, from that domain and also variables that form combinations with $u_{j_1},u_{j_2},\dots$  which are prohibited by conservation laws. This strategy  ensures that after $n$ steps one of the terms of determinant turns into~1. 
Since all the rest of the terms in the determinant are non negative, the $\textstyle{\det_{j_1,j_2,\dots,j_n}} v(u_1,\dots,u_s)$ is always positive in the whole integration domain and integrand itself has no divergences.

As an example, consider the diagram on Fig.~\ref{pic_eye}.
According to~\eqref{J_4} this graph corresponds to the following integral in Feynman representation:
\be
\label{sd_J_2_eye}
J_{2}^{(4)} =\frac{\Gamma \left(4- d\right)m^{2(d-4)}} {\left(4\pi \right)^{d}
            }\ I_2^{(4)},\quad\text{where}\quad
    I_2^{(4)}=\int \limits _{0}^{1}\int \limits _{0}^{1} \int \limits _{0}^{1}du_{1} du_{2} du_{3}
    \frac{u_{1} \delta \left(\sum \limits _{i=1}^{3}u_{i} -1\right)}
          {\left(\det v\right)^{d/2}}, \quad
          \det v = u_1 u_2 + u_2 u_3 + u_1 u_3.
\ee 

\begin{figure}[h]
\begin{tikzpicture}[node distance=2cm]
\coordinate[vertex] (v1); 
\coordinate[vertex, below=of v1] (v2);
\coordinate(x1) at ($ (v1)!.5!(v2) $);
\coordinate[vertex, left=of x1] (v3);

\arcll{v1}{v2}{35}{2};
\arcrl{v1}{v2}{35}{3};
\arcrl{v3}{v1}{0}{1};
\arcrl{v3}{v2}{0}{1};
\draw (v3) -- ($ (v3)!.15!180:(v1) $);
\draw (v3) -- ($ (v3)!.15!180:(v2) $);
\draw (v2) -- ($ (v2)!.15!180:(v3) $);
\draw (v1) -- ($ (v1)!.15!180:(v3) $);
\end{tikzpicture}
\caption{Two loop vertex diagram ($ee12|22|e|$). Number on the edge denotes the Feynman parameter corresponding to this edge.}
\label{pic_eye}
\end{figure}


 Following the described strategy,  integral $I_2^{(4)}$ can be expressed as a sum of the following six sectors: $[u_1,u_2]$, $[u_1,u_3]$, $[u_2,u_1]$, $[u_2,u_3]$, $[u_3,u_1]$, $[u_3,u_2]$. As an example let us consider the sector $[u_1,u_2]$. 
At the first step of decomposition main variable in this sector is $u_1$.
Decomposition domain includes all three variables $\{u_1, u_2, u_3\}$. Following \eqref{sd_1},\eqref{sd_2} one needs to stretch variables $u_2$ and $u_3$ in the following way: $u_2 \to u_1 u_2$, and $u_3 \to u_1 u_3$.
Denoting operation of decomposition over $u_1$ as $\hat{S_1}$, we obtain:
$$
\sHat_1 I_2^{(4)}=\int \limits _{0}^{1}\int \limits _{0}^{1} \int \limits _{0}^{1}du_{1} du_{2} du_{3}
    \frac{u_{1}^{3-d} \ 
    \delta \left(u_1(1+u_2+u_3) -1 \right)}
          {\left(u_2 + u_2 u_3 + u_3\right)^{d/2}}.
$$

We can eliminate $\delta$-function from integral by integrating over $u_1$.
Taking into account
$$
\delta[u_1(1+u_2+u_3)-1]=\frac{1}{1+u_2+u_3}
	\delta\left(u_1-\frac{1}{1+u_2+u_3}\right),
$$
and $0<1/(1+u_2+u_3)<1$,
we obtain
$$
\sHat_1 I_2^{(4)}=\int \limits _{0}^{1} \int \limits _{0}^{1} du_{2} du_{3}\ 
    \frac{1}
          {\left(u_2 + u_2 u_3 + u_3\right)^{d/2}(1+u_2+u_3)^{4-d}}.
$$
At the second step of decomposition, the main variable is $u_2$, decomposition domain is $\{u_2,u_3\}$. According to \eqref{sd_1},\eqref{sd_2} we have:

\be
\label{sd_sec_12}
\sHat_{12}I_2^{(4)}=\int \limits _{0}^{1} \int \limits _{0}^{1} du_{2} du_{3}\ 
    \frac{u_2^{1-d/2}}
          {\left(1 + u_2 u_3 + u_3\right)^{d/2}(1+u_2 u_3+u_3)^{4-d}}.
\ee
In the case $d=2$ this integral can be easily calculated numerically. 

Results for the rest of sectors are given below.
Variables $u_2$ and $u_3$ are included in $I_2^{(4)}$ symmetrically, and the contribution of the sector $[u_1,u_3]$ is equal to the contribution of the sector $[u_1,u_2]$, and is given in~\eqref{sd_sec_12}. For the same reason pairwise equal are $[u_2,u_1]$ and $[u_3,u_1]$, $[u_2,u_3]$ and $[u_3,u_2]$. After calculations similar to the above-stated, we obtain: 

\be
\label{sd_sec_21}
\begin{aligned}
\sHat_{21}I_2^{(4)}=\sHat_{31}I_2^{(4)}=
	\int \limits _{0}^{1} \int \limits _{0}^{1} du_{1} du_{3}\ 
    \frac{u_1^{2-d/2}}
          {\left(1 + u_1 u_3 + u_3\right)^{d/2}(1+u_1 u_3+u_1)^{3-d}},\\
\sHat_{23}I_2^{(4)}=\sHat_{32}I_2^{(4)}=
	\int \limits _{0}^{1} \int \limits _{0}^{1} du_{1} du_{3}\ 
    \frac{u_1\,u_3^{2-d/2}}
          {\left(1 + u_1 u_3 + u_1\right)^{d/2}(1+u_1 u_3+u_3)^{3-d}}.          
\end{aligned}
\ee
One can note that the reduction in the number of independent sectors was achieved (twice) due to the symmetry of the original diagram. Let us examine this question in more detail. 

\subsection{Using symmetries}

For the identification of equivalent sectors it is convenient to associate each edge of the diagram  with "multi-index", that reflects the "role" of this edge at each step of decomposition. For example, the index $(m,s,n)$ (the order of letters is significant!) means that some edge at the first step of decomposition was the "main", at the second step -- "secondary", and at the third one -- it was not the part of decomposition space ("neutral"). Constructed in such a way \textit{ diagram with multi-indices} identifies sector uniquely; in other words, according to the diagram with multi-indices one can reconstruct the sector. 
Sectors will be equivalent, if corresponding diagrams with multi-indices are isomorphic.

The consideration of symmetries using a multi-index is suitable for any strategy of decomposition. As noted above, in our scheme assignment of the sector is completely determined by the indication of main variable at each step of decomposition. Therefore, the multi-index assignment can be replaced by such indication.

Consider as an example the diagram depicted in Fig.\ref{3loop2leg} as a completely labeled graph (with both vertices and edges labeled). 
Denoting with Roman numerals the order of selection of the main edges on each of the three steps of decomposition, we obtain 6 combinatorially different diagrams drawn in  Fig.\ref{symm_colored_diags}:
\begin{figure}[h]
\begin{tikzpicture}[node distance=2cm] 
\coordinate[vertex] (v1); 
\coordinate[vertex, right=of v1] (v2);
\coordinate[vertex, right=of v2] (v3);
 
\coordinate[right=of v3] (xe2);
\coordinate[left=of v1] (xe1);
\coordinate(e2) at ($ (v3)!.15!(xe2) $);
\coordinate(e1) at ($ (v1)!.15!(xe1) $);

\arcrl{v1}{v3}{55}{III};

\arcl{v2}{v1}{35};
\arcrl{v2}{v1}{35}{I};

\arcl{v3}{v2}{35};
\arcrl{v3}{v2}{35}{II};

\draw (v1) -- (e1);
\draw (v3) -- (e2);

\end{tikzpicture}\quad
\begin{tikzpicture}[node distance=2cm]
\coordinate[vertex] (v1); 
\coordinate[vertex, right=of v1] (v2);
\coordinate[vertex, right=of v2] (v3);
 
\coordinate[right=of v3] (xe2);
\coordinate[left=of v1] (xe1);
\coordinate(e2) at ($ (v3)!.15!(xe2) $);
\coordinate(e1) at ($ (v1)!.15!(xe1) $);

\arcrl{v1}{v3}{55}{II};

\arcl{v2}{v1}{35};
\arcrl{v2}{v1}{35}{I};

\arcl{v3}{v2}{35};
\arcrl{v3}{v2}{35}{III};

\draw (v1) -- (e1);
\draw (v3) -- (e2);

\end{tikzpicture}
\quad
\begin{tikzpicture}[node distance=2cm]
\coordinate[vertex] (v1); 
\coordinate[vertex, right=of v1] (v2);
\coordinate[vertex, right=of v2] (v3);
 
\coordinate[right=of v3] (xe2);
\coordinate[left=of v1] (xe1);
\coordinate(e2) at ($ (v3)!.15!(xe2) $);
\coordinate(e1) at ($ (v1)!.15!(xe1) $);

\arcrl{v1}{v3}{55}{I};

\arcl{v2}{v1}{35};
\arcrl{v2}{v1}{35}{II};

\arcl{v3}{v2}{35};
\arcrl{v3}{v2}{35}{III};

\draw (v1) -- (e1);
\draw (v3) -- (e2);

\end{tikzpicture}

\begin{tikzpicture}[node distance=2cm]
\coordinate[vertex] (v1); 
\coordinate[vertex, right=of v1] (v2);
\coordinate[vertex, right=of v2] (v3);
 
\coordinate[right=of v3] (xe2);
\coordinate[left=of v1] (xe1);
\coordinate(e2) at ($ (v3)!.15!(xe2) $);
\coordinate(e1) at ($ (v1)!.15!(xe1) $);

\arcr{v1}{v3}{55};

\arcll{v2}{v1}{35}{II};
\arcrl{v2}{v1}{35}{I};

\arcl{v3}{v2}{35};
\arcrl{v3}{v2}{35}{III};

\draw (v1) -- (e1);
\draw (v3) -- (e2);

\end{tikzpicture}\quad
\begin{tikzpicture}[node distance=2cm]
\coordinate[vertex] (v1); 
\coordinate[vertex, right=of v1] (v2);
\coordinate[vertex, right=of v2] (v3);
 
\coordinate[right=of v3] (xe2);
\coordinate[left=of v1] (xe1);
\coordinate(e2) at ($ (v3)!.15!(xe2) $);
\coordinate(e1) at ($ (v1)!.15!(xe1) $);

\arcr{v1}{v3}{55};

\arcll{v2}{v1}{35}{III};
\arcrl{v2}{v1}{35}{I};

\arcl{v3}{v2}{35};
\arcrl{v3}{v2}{35}{II};

\draw (v1) -- (e1);
\draw (v3) -- (e2);

\end{tikzpicture}
\quad
\begin{tikzpicture}[node distance=2cm]
\coordinate[vertex] (v1); 
\coordinate[vertex, right=of v1] (v2);
\coordinate[vertex, right=of v2] (v3);
 
\coordinate[right=of v3] (xe2);
\coordinate[left=of v1] (xe1);
\coordinate(e2) at ($ (v3)!.15!(xe2) $);
\coordinate(e1) at ($ (v1)!.15!(xe1) $);

\arcr{v1}{v3}{55};

\arcl{v2}{v1}{35};
\arcrl{v2}{v1}{35}{I};

\arcll{v3}{v2}{35}{III};
\arcrl{v3}{v2}{35}{II};

\draw (v1) -- (e1);
\draw (v3) -- (e2);

\end{tikzpicture}
\caption{Graph representation for nonequivalent sectors for graph $e112|22|e|$. Roman numerals mark the edges whose Feynman parameter is used as main variable on the corresponding decomposition step.}
\label{symm_colored_diags}
\end{figure}

For every diagram from~Fig.\ref{symm_colored_diags} there are 7 more isomorphic diagrams that give equal contribution. Isomorphic sectors for the first diagram in Fig.\ref{symm_colored_diags} are drawn in Fig.\ref{symm_colored_diags_isom}.
\begin{figure}[h]
\begin{tikzpicture}[node distance=1.5cm]
\coordinate[vertex] (v1); 
\coordinate[vertex, right=of v1] (v2);
\coordinate[vertex, right=of v2] (v3);
 
\coordinate[right=of v3] (xe2);
\coordinate[left=of v1] (xe1);
\coordinate(e2) at ($ (v3)!.15!(xe2) $);
\coordinate(e1) at ($ (v1)!.15!(xe1) $);

\arcrl{v1}{v3}{55}{III};

\arcl{v2}{v1}{35};
\arcrl{v2}{v1}{35}{I};

\arcl{v3}{v2}{35};
\arcrl{v3}{v2}{35}{II};

\draw (v1) -- (e1);
\draw (v3) -- (e2);

\end{tikzpicture}
\ 
\begin{tikzpicture}[node distance=1.5cm]
\coordinate[vertex] (v1); 
\coordinate[vertex, right=of v1] (v2);
\coordinate[vertex, right=of v2] (v3);
 
\coordinate[right=of v3] (xe2);
\coordinate[left=of v1] (xe1);
\coordinate(e2) at ($ (v3)!.15!(xe2) $);
\coordinate(e1) at ($ (v1)!.15!(xe1) $);

\arcrl{v1}{v3}{55}{III};

\arcl{v2}{v1}{35};
\arcrl{v2}{v1}{35}{I};

\arcll{v3}{v2}{35}{II};
\arcr{v3}{v2}{35};

\draw (v1) -- (e1);
\draw (v3) -- (e2);

\end{tikzpicture}
\ 
\begin{tikzpicture}[node distance=1.5cm]
\coordinate[vertex] (v1); 
\coordinate[vertex, right=of v1] (v2);
\coordinate[vertex, right=of v2] (v3);
 
\coordinate[right=of v3] (xe2);
\coordinate[left=of v1] (xe1);
\coordinate(e2) at ($ (v3)!.15!(xe2) $);
\coordinate(e1) at ($ (v1)!.15!(xe1) $);

\arcrl{v1}{v3}{55}{III};

\arcll{v2}{v1}{35}{I};
\arcr{v2}{v1}{35};

\arcll{v3}{v2}{35}{II};
\arcr{v3}{v2}{35};

\draw (v1) -- (e1);
\draw (v3) -- (e2);

\end{tikzpicture}
\ 
\begin{tikzpicture}[node distance=1.5cm]
\coordinate[vertex] (v1); 
\coordinate[vertex, right=of v1] (v2);
\coordinate[vertex, right=of v2] (v3);
 
\coordinate[right=of v3] (xe2);
\coordinate[left=of v1] (xe1);
\coordinate(e2) at ($ (v3)!.15!(xe2) $);
\coordinate(e1) at ($ (v1)!.15!(xe1) $);

\arcrl{v1}{v3}{55}{III};

\arcll{v2}{v1}{35}{I};
\arcr{v2}{v1}{35};

\arcl{v3}{v2}{35};
\arcrl{v3}{v2}{35}{II};

\draw (v1) -- (e1);
\draw (v3) -- (e2);

\end{tikzpicture}

\begin{tikzpicture}[node distance=1.5cm]
\coordinate[vertex] (v1); 
\coordinate[vertex, right=of v1] (v2);
\coordinate[vertex, right=of v2] (v3);
 
\coordinate[right=of v3] (xe2);
\coordinate[left=of v1] (xe1);
\coordinate(e2) at ($ (v3)!.15!(xe2) $);
\coordinate(e1) at ($ (v1)!.15!(xe1) $);

\arcrl{v1}{v3}{55}{III};

\arcl{v2}{v1}{35};
\arcrl{v2}{v1}{35}{II};

\arcl{v3}{v2}{35};
\arcrl{v3}{v2}{35}{I};

\draw (v1) -- (e1);
\draw (v3) -- (e2);

\end{tikzpicture}
\ 
\begin{tikzpicture}[node distance=1.5cm]
\coordinate[vertex] (v1); 
\coordinate[vertex, right=of v1] (v2);
\coordinate[vertex, right=of v2] (v3);
 
\coordinate[right=of v3] (xe2);
\coordinate[left=of v1] (xe1);
\coordinate(e2) at ($ (v3)!.15!(xe2) $);
\coordinate(e1) at ($ (v1)!.15!(xe1) $);

\arcrl{v1}{v3}{55}{III};

\arcl{v2}{v1}{35};
\arcrl{v2}{v1}{35}{II};

\arcll{v3}{v2}{35}{I};
\arcr{v3}{v2}{35};

\draw (v1) -- (e1);
\draw (v3) -- (e2);

\end{tikzpicture}
\ 
\begin{tikzpicture}[node distance=1.5cm]
\coordinate[vertex] (v1); 
\coordinate[vertex, right=of v1] (v2);
\coordinate[vertex, right=of v2] (v3);
 
\coordinate[right=of v3] (xe2);
\coordinate[left=of v1] (xe1);
\coordinate(e2) at ($ (v3)!.15!(xe2) $);
\coordinate(e1) at ($ (v1)!.15!(xe1) $);

\arcrl{v1}{v3}{55}{III};

\arcll{v2}{v1}{35}{II};
\arcr{v2}{v1}{35};

\arcll{v3}{v2}{35}{I};
\arcr{v3}{v2}{35};

\draw (v1) -- (e1);
\draw (v3) -- (e2);

\end{tikzpicture}
\ 
\begin{tikzpicture}[node distance=1.5cm]
\coordinate[vertex] (v1); 
\coordinate[vertex, right=of v1] (v2);
\coordinate[vertex, right=of v2] (v3);
 
\coordinate[right=of v3] (xe2);
\coordinate[left=of v1] (xe1);
\coordinate(e2) at ($ (v3)!.15!(xe2) $);
\coordinate(e1) at ($ (v1)!.15!(xe1) $);

\arcrl{v1}{v3}{55}{III};

\arcll{v2}{v1}{35}{II};
\arcr{v2}{v1}{35};

\arcl{v3}{v2}{35};
\arcrl{v3}{v2}{35}{I};

\draw (v1) -- (e1);
\draw (v3) -- (e2);

\end{tikzpicture}
\caption{Graph representation for sectors isomorphic to the first sector in Fig.\ref{symm_colored_diags} }
\label{symm_colored_diags_isom}
\end{figure}
Therefore, instead of calculating all the 48 sectors for the diagram on Fig.\ref{3loop2leg}, it is sufficient to calculate only 6 independent sectors with weight factor  8. 

For the graph under consideration it is easy to find isomorphic graphs, but in general case it becomes a non-trivial problem. To find isomorphic graphs with labeled edges we use GraphState package~\cite{GraphState}.

\subsection{$R'$-operation in the Feynman representation}
\label{R-prime}

Above (see Section \ref{sec_feynman}) we noted, that for any diagram its Feynman representation can be written directly (bypassing momentum representation). In the same fashion, subtraction operation~\eqref{1-K} can be formulated in terms of Feynman parameters.

\begin{wrapfigure}{r}{0.35\textwidth}
\begin{tikzpicture}[node distance=1.6cm]
\coordinate[vertex] (v1); 
\coordinate[vertex,above right=of v1] (v3);

\coordinate[vertex, right=of v3] (v4);
\coordinate[vertex, below right=of v4] (v2);

\arcls{v2}{v1}{35}{1};
\arcls{v1}{v3}{25}{1};
\arcls{v2}{v4}{-25}{1};
\arcls{v3}{v4}{75}{2};
\arcls{v3}{v4}{0}{3};
\arcls{v3}{v4}{-75}{4};

\coordinate[above left=of v1] (xe2);
\coordinate[below left=of v1] (xe1);
\coordinate(e2) at ($ (v1)!.25!(xe2) $);
\coordinate(e1) at ($ (v1)!.25!(xe1) $);
\draw (v1) -- (e1);
\draw (v1) -- (e2);

\coordinate[above right=of v2] (xe3);
\coordinate[below right=of v2] (xe4);
\coordinate(e3) at ($ (v2)!.25!(xe3) $);
\coordinate(e4) at ($ (v2)!.25!(xe4) $);
\draw (v2) -- (e3);
\draw (v2) -- (e4);

\end{tikzpicture}
	\caption{Vertex diagram ($ee12|ee3|333||$) with quadratically divergent subgraph }
    \label{pic:bubble-sunset}
\end{wrapfigure}
 It can be shown, that the introduction of the stretching parameters $a$ of momenta flowing  into subgraphs (according to the equation~\eqref{1-K}) leads to the following modification of  $\det v$ and $Q$ in the Feynman representation~\eqref{J_4},~\eqref{J_2}. Assume that parameter $a$ corresponds to an $\cal L$-loop subgraph, and $s$ is a number of Feynman parameters belonging to this subgraph in a term $\det v$ or $Q$. Then, for $s{>}{\cal L}$ we should multiply this term with $a^{s-\cal L} $. Consider as an example the diagram in Fig.~\ref{pic:bubble-sunset}.

For this diagram, in accordance with the rules above:

$$\det v=u_{1} u_{2} u_{3} +u_{1} u_{2} u_{4} +u_{1} u_{3} u_{4} +a\, u_{2} u_{3} u_{4}.$$ 

Using~\eqref{J_4}, one can find
\be
\label{sd_example}
\begin{aligned}
&R_{2}'J_{3}^{(4)} =
	\frac{\Gamma \left(3(2-d/2)\right)m^{3(d-4)}} {2\left(4\pi \right)^{3d/2}}
    	\int\limits_{0}^{1}du_{1} \dots \int \limits_{0}^{1} du_{4} \int \limits_{0}^{1}da\,
        \frac12 \left(1-a\right)\, \partial_{a}^{2}   
        \,\frac{\delta \left(\sum \limits_{i=1}^{4}u_{i}  -1\right) u_{1}^{2}}{\left(\det v\right)^{d/2} } =\\
        & = \frac{d(d+2)\Gamma \left(3(2-d/2)\right)m^{3(d-4)}}
        {16\left(4\pi \right)^{3d/2}}
        	\int \limits_{0}^{1}du_{1}\dots\int \limits_{0}^{1} du_{4} \int \limits_{0}^{1}da \left(1-a\right)
            \frac{\delta \left(\sum \limits_{i=1}^{4}u_{i} -1\right) \left(u_{1} u_{2} u_{3} u_{4} \right)^{2}} {\left(u_1 u_2 u_3 +u_1 u_2 u_4 +u_1 u_3 u_4 +a u_2 u_3 u_4 \right)^{d/2+2}} .
\end{aligned}
\ee

For this integral the sector decomposition strategy described in section~\ref{sd_scheme} cannot be applied directly because of the presence of stretching parameter $a$, but it is possible to modify decomposition strategy for such a case  (see Appendix~\ref{appendix_sd} for details).

As an example of $R'_4$ operation \eqref{L-operation}, consider the diagram on Fig.\ref{4loop4leg}:
\begin{equation}
\begin{aligned}
R'_4\;\begin{matrix}\\[5pt] 
\begin{tikzpicture}[node distance=0.6cm]
\coordinate[vertex] (v1); 
\coordinate[vertex, right=of v1] (v2);
\coordinate[vertex, right=of v2] (v3);
\coordinate[vertex, right=of v3] (v4);
\coordinate(x1) at ($ (v2)!.5!(v3) $);
\coordinate[vertex] (v5) at ($ (x1)!1.4!270:(v3) $);
 
\coordinate[right=of v4] (xe2);
\coordinate[left=of v1] (xe1);
\coordinate[below left=of v5] (xe3);
\coordinate[below right=of v5] (xe4);
\coordinate(e2) at ($ (v4)!.25!(xe2) $);
\coordinate(e1) at ($ (v1)!.25!(xe1) $);
\coordinate(e3) at ($ (v5)!.25!(xe3) $);
\coordinate(e4) at ($ (v5)!.25!(xe4) $);

\arcl{v2}{v1}{45};
\arcr{v2}{v1}{45};

\arcl{v3}{v2}{45};
\arcr{v3}{v2}{45};

\arcl{v4}{v3}{45};
\arcr{v4}{v3}{45};

\draw (v1) -- (e1);
\draw (v4) -- (e2);
\draw (v5) -- (e3);
\draw (v5) -- (e4);
\arcl{v5}{v1}{25};
\arcr{v5}{v4}{25};

\end{tikzpicture}
\end{matrix}\;=\;
\begin{matrix}\\[5pt]
\begin{tikzpicture}[node distance=0.6cm]
\coordinate[vertex] (v1); 
\coordinate[vertex, right=of v1] (v2);
\coordinate[vertex, right=of v2] (v3);
\coordinate[vertex, right=of v3] (v4);
\coordinate(x1) at ($ (v2)!.5!(v3) $);
\coordinate[vertex] (v5) at ($ (x1)!1.4!270:(v3) $);
 
\coordinate[right=of v4] (xe2);
\coordinate[left=of v1] (xe1);
\coordinate[below left=of v5] (xe3);
\coordinate[below right=of v5] (xe4);
\coordinate(e2) at ($ (v4)!.25!(xe2) $);
\coordinate(e1) at ($ (v1)!.25!(xe1) $);
\coordinate(e3) at ($ (v5)!.25!(xe3) $);
\coordinate(e4) at ($ (v5)!.25!(xe4) $);

\arcl{v2}{v1}{45};
\arcr{v2}{v1}{45};

\arcl{v3}{v2}{45};
\arcr{v3}{v2}{45};

\arcl{v4}{v3}{45};
\arcr{v4}{v3}{45};

\draw (v1) -- (e1);
\draw (v4) -- (e2);
\draw (v5) -- (e3);
\draw (v5) -- (e4);
\arcl{v5}{v1}{25};
\arcr{v5}{v4}{25};
\end{tikzpicture}
\end{matrix}\; - \;
\begin{matrix}\\[3pt]
\begin{tikzpicture}[node distance=0.6cm]
\coordinate[vertex] (v1); 
\coordinate[vertex, right=of v1] (v2);
\coordinate[vertex, right=of v2] (v3);
\coordinate[vertex, right=of v3] (v4);
\coordinate(x1) at ($ (v2)!.5!(v3) $);
\coordinate[vertex] (v5) at ($ (x1)!1.4!270:(v3) $);
 
\coordinate[right=of v4] (xe2);
\coordinate[left=of v1] (xe1);
\coordinate[below left=of v5] (xe3);
\coordinate[below right=of v5] (xe4);
\coordinate(e2) at ($ (v4)!.25!(xe2) $);
\coordinate(e1) at ($ (v1)!.25!(xe1) $);
\coordinate(e3) at ($ (v5)!.25!(xe3) $);
\coordinate(e4) at ($ (v5)!.25!(xe4) $);

\arcl{v2}{v1}{55};
\arcr{v2}{v1}{55};

\arcl{v3}{v2}{55};
\arcr{v3}{v2}{55};

\arcl{v4}{v3}{55};
\arcr{v4}{v3}{55};

\draw (v1) -- (e1);
\draw (v4) -- (e2);
\draw (v5) -- (e3);
\draw (v5) -- (e4);
\arcl{v5}{v1}{25};
\arcr{v5}{v4}{25};

\draw[dashed] let \p1 = ($ (v1)-(v2) $),
              \n1 = {veclen(\x1,\y1)} 
              in circle [at=($ (v1)!.5!(v2) $), x radius=\n1*0.6, y radius=\n1*0.4];
\end{tikzpicture}
\end{matrix}\; -\;
\begin{matrix}\\[3pt]
\begin{tikzpicture}[node distance=0.6cm]
\coordinate[vertex] (v1); 
\coordinate[vertex, right=of v1] (v2);
\coordinate[vertex, right=of v2] (v3);
\coordinate[vertex, right=of v3] (v4);
\coordinate(x1) at ($ (v2)!.5!(v3) $);
\coordinate[vertex] (v5) at ($ (x1)!1.4!270:(v3) $);
 
\coordinate[right=of v4] (xe2);
\coordinate[left=of v1] (xe1);
\coordinate[below left=of v5] (xe3);
\coordinate[below right=of v5] (xe4);
\coordinate(e2) at ($ (v4)!.25!(xe2) $);
\coordinate(e1) at ($ (v1)!.25!(xe1) $);
\coordinate(e3) at ($ (v5)!.25!(xe3) $);
\coordinate(e4) at ($ (v5)!.25!(xe4) $);

\arcl{v2}{v1}{55};
\arcr{v2}{v1}{55};

\arcl{v3}{v2}{55};
\arcr{v3}{v2}{55};

\arcl{v4}{v3}{55};
\arcr{v4}{v3}{55};

\draw (v1) -- (e1);
\draw (v4) -- (e2);
\draw (v5) -- (e3);
\draw (v5) -- (e4);
\arcl{v5}{v1}{25};
\arcr{v5}{v4}{25};

\draw[dashed] let \p1 = ($ (v2)-(v3) $),
              \n1 = {veclen(\x1,\y1)} 
              in circle [at=($ (v2)!.5!(v3) $), x radius=\n1*0.6, y radius=\n1*0.4];
\end{tikzpicture}
\end{matrix}\; -\;
\begin{matrix}\\[3pt]
\begin{tikzpicture}[node distance=0.6cm]
\coordinate[vertex] (v1); 
\coordinate[vertex, right=of v1] (v2);
\coordinate[vertex, right=of v2] (v3);
\coordinate[vertex, right=of v3] (v4);
\coordinate(x1) at ($ (v2)!.5!(v3) $);
\coordinate[vertex] (v5) at ($ (x1)!1.4!270:(v3) $);
 
\coordinate[right=of v4] (xe2);
\coordinate[left=of v1] (xe1);
\coordinate[below left=of v5] (xe3);
\coordinate[below right=of v5] (xe4);
\coordinate(e2) at ($ (v4)!.25!(xe2) $);
\coordinate(e1) at ($ (v1)!.25!(xe1) $);
\coordinate(e3) at ($ (v5)!.25!(xe3) $);
\coordinate(e4) at ($ (v5)!.25!(xe4) $);

\arcl{v2}{v1}{55};
\arcr{v2}{v1}{55};

\arcl{v3}{v2}{55};
\arcr{v3}{v2}{55};

\arcl{v4}{v3}{55};
\arcr{v4}{v3}{55};

\draw (v1) -- (e1);
\draw (v4) -- (e2);
\draw (v5) -- (e3);
\draw (v5) -- (e4);
\arcl{v5}{v1}{25};
\arcr{v5}{v4}{25};

\draw[dashed] let \p1 = ($ (v3)-(v4) $),
              \n1 = {veclen(\x1,\y1)} 
              in circle [at=($ (v3)!.5!(v4) $), x radius=\n1*0.6, y radius=\n1*0.4];
\end{tikzpicture}
\end{matrix}\; -\\ 
- \;
\begin{matrix}\\[1pt]
\begin{tikzpicture}[node distance=0.6cm]
\coordinate[vertex] (v1); 
\coordinate[vertex, right=of v1] (v2);
\coordinate[vertex, right=of v2] (v3);
\coordinate[vertex, right=of v3] (v4);
\coordinate(x1) at ($ (v2)!.5!(v3) $);
\coordinate[vertex] (v5) at ($ (x1)!1.4!270:(v3) $);
 
\coordinate[right=of v4] (xe2);
\coordinate[left=of v1] (xe1);
\coordinate[below left=of v5] (xe3);
\coordinate[below right=of v5] (xe4);
\coordinate(e2) at ($ (v4)!.25!(xe2) $);
\coordinate(e1) at ($ (v1)!.25!(xe1) $);
\coordinate(e3) at ($ (v5)!.25!(xe3) $);
\coordinate(e4) at ($ (v5)!.25!(xe4) $);

\arcl{v2}{v1}{55};
\arcr{v2}{v1}{55};

\arcl{v3}{v2}{55};
\arcr{v3}{v2}{55};

\arcl{v4}{v3}{55};
\arcr{v4}{v3}{55};

\draw (v1) -- (e1);
\draw (v4) -- (e2);
\draw (v5) -- (e3);
\draw (v5) -- (e4);
\arcl{v5}{v1}{25};
\arcr{v5}{v4}{25};

\draw[dashed] let \p1 = ($ (v1)-(v4) $),
              \n1 = {veclen(\x1,\y1)} 
              in circle [at=($ (v1)!.5!(v4) $), x radius=\n1*0.54, y radius=\n1*0.16];
\end{tikzpicture}
\end{matrix}\;-\;
\begin{matrix}\\[1pt]
\begin{tikzpicture}[node distance=0.6cm]
\coordinate[vertex] (v1); 
\coordinate[vertex, right=of v1] (v2);
\coordinate[vertex, right=of v2] (v3);
\coordinate[vertex, right=of v3] (v4);
\coordinate(x1) at ($ (v2)!.5!(v3) $);
\coordinate[vertex] (v5) at ($ (x1)!1.4!270:(v3) $);
 
\coordinate[right=of v4] (xe2);
\coordinate[left=of v1] (xe1);
\coordinate[below left=of v5] (xe3);
\coordinate[below right=of v5] (xe4);
\coordinate(e2) at ($ (v4)!.25!(xe2) $);
\coordinate(e1) at ($ (v1)!.25!(xe1) $);
\coordinate(e3) at ($ (v5)!.25!(xe3) $);
\coordinate(e4) at ($ (v5)!.25!(xe4) $);

\arcl{v2}{v1}{55};
\arcr{v2}{v1}{55};

\arcl{v3}{v2}{55};
\arcr{v3}{v2}{55};

\arcl{v4}{v3}{55};
\arcr{v4}{v3}{55};

\draw (v1) -- (e1);
\draw (v4) -- (e2);
\draw (v5) -- (e3);
\draw (v5) -- (e4);
\arcl{v5}{v1}{25};
\arcr{v5}{v4}{25};

\draw[dashed] let \p1 = ($ (v1)-(v3) $),
              \n1 = {veclen(\x1,\y1)} 
              in circle [at=($ (v1)!.5!(v3) $), x radius=\n1*0.55, y radius=\n1*0.25];
\end{tikzpicture}
\end{matrix}\; - \;
\begin{matrix}\\[1pt]
\begin{tikzpicture}[node distance=0.6cm]
\coordinate[vertex] (v1); 
\coordinate[vertex, right=of v1] (v2);
\coordinate[vertex, right=of v2] (v3);
\coordinate[vertex, right=of v3] (v4);
\coordinate(x1) at ($ (v2)!.5!(v3) $);
\coordinate[vertex] (v5) at ($ (x1)!1.4!270:(v3) $);
 
\coordinate[right=of v4] (xe2);
\coordinate[left=of v1] (xe1);
\coordinate[below left=of v5] (xe3);
\coordinate[below right=of v5] (xe4);
\coordinate(e2) at ($ (v4)!.25!(xe2) $);
\coordinate(e1) at ($ (v1)!.25!(xe1) $);
\coordinate(e3) at ($ (v5)!.25!(xe3) $);
\coordinate(e4) at ($ (v5)!.25!(xe4) $);

\arcl{v2}{v1}{55};
\arcr{v2}{v1}{55};

\arcl{v3}{v2}{55};
\arcr{v3}{v2}{55};

\arcl{v4}{v3}{55};
\arcr{v4}{v3}{55};

\draw (v1) -- (e1);
\draw (v4) -- (e2);
\draw (v5) -- (e3);
\draw (v5) -- (e4);
\arcl{v5}{v1}{25};
\arcr{v5}{v4}{25};

\draw[dashed] let \p1 = ($ (v2)-(v4) $),
              \n1 = {veclen(\x1,\y1)} 
              in circle [at=($ (v2)!.5!(v4) $), x radius=\n1*0.55, y radius=\n1*0.25];
\end{tikzpicture}
\end{matrix}\; +\;
\begin{matrix}\\[3pt]
\begin{tikzpicture}[node distance=0.6cm]
\coordinate[vertex] (v1); 
\coordinate[vertex, right=of v1] (v2);
\coordinate[vertex, right=of v2] (v3);
\coordinate[vertex, right=of v3] (v4);
\coordinate(x1) at ($ (v2)!.5!(v3) $);
\coordinate[vertex] (v5) at ($ (x1)!1.4!270:(v3) $);
 
\coordinate[right=of v4] (xe2);
\coordinate[left=of v1] (xe1);
\coordinate[below left=of v5] (xe3);
\coordinate[below right=of v5] (xe4);
\coordinate(e2) at ($ (v4)!.25!(xe2) $);
\coordinate(e1) at ($ (v1)!.25!(xe1) $);
\coordinate(e3) at ($ (v5)!.25!(xe3) $);
\coordinate(e4) at ($ (v5)!.25!(xe4) $);

\arcl{v2}{v1}{55};
\arcr{v2}{v1}{55};

\arcl{v3}{v2}{55};
\arcr{v3}{v2}{55};

\arcl{v4}{v3}{55};
\arcr{v4}{v3}{55};

\draw (v1) -- (e1);
\draw (v4) -- (e2);
\draw (v5) -- (e3);
\draw (v5) -- (e4);
\arcl{v5}{v1}{25};
\arcr{v5}{v4}{25};

\draw[dashed] let \p1 = ($ (v1)-(v2) $),
              \n1 = {veclen(\x1,\y1)} 
              in circle [at=($ (v1)!.5!(v2) $), x radius=\n1*0.6, y radius=\n1*0.4];
\draw[dashed] let \p1 = ($ (v3)-(v4) $),
              \n1 = {veclen(\x1,\y1)} 
              in circle [at=($ (v3)!.5!(v4) $), x radius=\n1*0.6, y radius=\n1*0.4];
\end{tikzpicture}
\end{matrix}\, .
\end{aligned}
\nonumber
\end{equation}

Here the dashed line around the subgraph denotes an action of counterterm operation $L$ (from  \eqref{L-operation}) on it. Note that all the terms with dashed lines consist of the factors calculated in the previous steps of the perturbation theory.

%% file: resummation.tex
All methods of calculating diagrams described in the preceding paragraphs have been implemented as the set of computer programs which includes:
\begin{itemize}
\item diagram generation, determination of the symmetry factors, subgraph identification, graph isomorphism detection;

\item calculation of the diagrams using sector decomposition technique (evaluation of the integrals is performed using \textsc{cuba} library~\cite{Hahn2007});
\item implementation of $R'$-operation;
\item calculation of $\beta$-function and anomalous dimensions.
\end{itemize}
The details of the software implementation see in~\cite{GraphState}.

With this software, all the diagrams with 4 legs up to 5 loops, and all the diagrams with 2 legs up to 6 loops were calculated numerically.
The results of calculation of individual diagrams are shown in the Appendix~\ref{appendix_diags}.

\subsection{Raw results}


The results of our calculations are as follows ($u=g/(4\pi)$):
\be
\label{beta_analytic}
	\begin{aligned}
	\frac{\beta(u,N)}{2}
= - u & \Big[\, 1 - \frac{N+8}{6} u + u^2(0.2870836262784(12)\,N + 1.324307020281(5))+\\
&- u^3%
(0.0231494256(15)\,N^2 + 0.690517865(27)\,N + 2.42766946(8))+\\
&+ u^4(6.85514(18)\times^{-5}\,N^3 + 0.13865554(5)\,N^2 + 2.0147798(4)\,N + 5.8573367(9)) -\\
&- u^5(-5.248(13)\times10^{-7}\,N^4 + 0.0103265(9)\,N^3 + 0.675546(15)\,N^2 + 6.84341(10)\,N +\\ 
&+ 17.22769(20)\,)\Big],\\
	\end{aligned}
\ee
 
\be
\label{eta_analytic}
\begin{aligned}
\gamma_2(u,N) = & (N+2)\,u^2%
	\Big[\,0.02547461027328(24)-0.0020225547250(11)\,(N+8)\,u+\\
   +&\  \big(-7.151583255(6)\times10^{-5}\,N^2 + 0.00327297797(25)\,N + 0.0160243885(11)\big)\,u^2-\\%
   -&\ \big (9.120353783(23)\times10^{-6}\,N^3 + 0.000117086(12)\,N^2 + 0.00719377(16)\,N +0.0274489(5)\big )\,u^3+\\%
   +&\ \big (-1.119698691(9)\times10^{-6}\,N^4 -2.3618(23)\times 10^{-5}\,N^3  +0.00142217(35)\,N^2 + \\ 
   +&\ 0.0259279(23)\,N + 0.079870(5)\big )\,u^4\Big].\\
\end{aligned}
\ee
This results  are in good agreement with the results of~\cite{Orlov2000, Nickel1978}. For ease of comparison, we give them one more time 
 in terms of the charge introduced in the article~\cite{Nickel1978}
($v(N) = u(N+8)/6 $), $\beta$-function is reproduced in the form similar to that used in~\cite{Orlov2000}:
\be
\label{beta_analytic_0}
	\begin{aligned}
	\frac{\beta(v,N)}{2}
= - v&\Big[\,1 - v + \frac{v^2}{(N+8)^2}\big(10.33501054602(4)\,N + 47.67505273011(17)\big)+\\
&- \frac{v^3%
}{(N+8)^3} \big(5.00027594(32)\,N^2 + 149.151859(6)\,N + 524.376603(17)\big)+\\
&+ \frac{v^4}{(N+8)^4}\big(0.0888427(23)\,N^3 + 179.69758(6)\,N^2 + 2611.1547(5)\,N + 7591.1084(11)\big)+\\
&- \frac{v^5}{(N+8)^5} \big(-0.004081(10)\,N^4 + 80.299(7)\,N^3 + 5253.04(11)\,N^2 + \\ &+ 53214.4(8)\,N + 133962.5(1.6)\big)\Big],\\
	\end{aligned}
\ee

\be
\label{eta_analytic_0}
\begin{aligned}
\gamma_2(v,N) = & v^2\frac{(N+2)}{(N+8)^2}%
	\Big[\,0.917085969838(9)-0.054608977575(29)\,v+\\
   +&\  (-0.09268451899(8)\,N^2 + 4.24177945(32)\,N + 20.7676074(15))\,\frac{v^2}{(N+8)^2}-\\
   -&\ (0.07091987102(18)\,N^3 + 0.91046(9)\,N^2 + 55.9387(12)\,N + 213.443(4))\,\frac{v^3}{(N+8)^3}+\\
   +&\ (-0.0522406621(4)\,N^4 - 1.1019(11)\,N^3  + 66.353(16)\,N^2 + 1209.69(11)\,N + 3726.43(23))\,\frac{v^4}{(N+8)^4}\Big].\\
\end{aligned}
\ee
For the case, when $N=1$ ($2D$ Ising model):

\be
\label{beta_raw}
\frac{\beta}{2} =  - v\Big[1 - v + 0.7161736206930(22)\,v^2 - 0.930766445(24)\,v^3 + 1.58238828(19)\,v^4 -  3.260178(30)\,v^5\Big].
\ee
In the result for $\gamma_2$ the first five terms of expansion are the same as in~\cite{Orlov2000}
and the sixth one is a new result obtained in this paper.
                        
\be
\label{eta_raw}
    \begin{aligned}
    \gamma_2\ =\  0.03396614703104(32) v^2 -\  &0.0020225547250(10) v^3 +  0.0113930966(7) v^4 \\
          - \ &0.01373587(21) v^5 + \mathbf{0.0282326(14)} v^6 .
    \end{aligned}
\ee
 
 The series \eqref{beta_raw} and \eqref{eta_raw} are asymptotic, so it is necessary to apply a resummation procedure to them.

\input{conform_Borel}

%% file: conform_Borel.tex
\subsection{Borel transform with conformal mapping}
As the resummation of results is only an auxiliary aim of this paper, we will not go into details of this question. Interested reader will find an exhaustive review and up to date bibliography on the methods of resummation in the report~\cite{Caliceti2007}.

For resummation of asymptotic series~\eqref{beta_raw},\eqref{eta_raw} we used the Borel transform with  conformal mapping of complex plane, which allows one to utilize the information about large-order behavior of the function.
Let us consider a quantity $A$ as a series in a parameter $v$:
$A(v) = \sum_k A_k v^k$ that possesses the following two properties: 
\begin{enumerate}
\item The first $n$ coefficients of this series are known from diagram calculation. 
\item Large-order behavior of this series is given by the following expression:
\be
\label{LOB}
A_k \simeq c(-a)^k\Gamma(k+b_0+1)\left[1+O\left(\frac1k\right)\right].
\ee
\end{enumerate}

The Borel transform with conformal mapping of complex plane allows to write a representation of $A(v)$ in a way that reflects both of this facts:
\begin{equation}
\label{borel_final}
A(v)=\sum_{k\ge1} U_k \int^{+\infty}_0 dt ~ t^{b} e^{-t} (w(v t))^k,
\end{equation}

where
\begin{equation}
\label{mapping}
w(x)=\frac{\sqrt{1+ax}-1}{\sqrt{1+ax}+1} \quad
\Leftrightarrow\quad x(w)=\frac{4w}{a(w-1)^2},
\end{equation}
and coefficients $U_k$ are chosen in a way that the expansion of~\eqref{borel_final} in $v$ would give the proper first $n$ coefficients. To reach this goal it is enough to truncate the series at $n^{\mathit{th}}$ term and choose $U_k$ in a form:

$$ 
   U_n=\sum^n_{m=1} \frac{A_m}{\Gamma(m+b+1)} (4/a)^m \binom{n{+}m{-}1}{n-m}, \quad n\ge1.
$$
The equation~\eqref{borel_final} predicts (see~\cite{Kazakov1979}) the large-order behavior of $A_k$ in a form:
\be
\label{LOB_Kazakov}%
A_k \simeq C(-a)^k\Gamma\left(k+b-\frac12\right)\left[1+O\left(\frac1k\right)\right].
\ee
For consistency between~\eqref{borel_final} and~\eqref{LOB_Kazakov} one should take:
\be
\label{resummation:b}
b = b_0 +3/2.
\ee
We also use \textit{shift of order K} (see part III.A in~\cite{Zinn1980}), it means that if the series to be resummed has number $K$ zero coefficients at the beginning, one should apply the resummation procedure to the function $A_{(K)}(v)$ defined from the following:
$$
A(v) = v^K A_{(K)}(v) = v^K (A_0^{(K)} + A_1^{(K)} v + \dots).
$$
Thus one should add $K$ to the right hand side of~\eqref{resummation:b}:
\be
b = b_0 +3/2 + K.
\ee
In our case $K=1$ for beta-function and $K=2$ for $\gamma_2$.

Asymptotic values of $a$ and $b_0$ in~\eqref{LOB} were calculated in~\cite{Brezin1978} (in terms of charge $v$):
$$
a = 0.238659217\ \frac{9}{N+8},\quad
b_0 = \left\{
\begin{array}{l}
2.5+N/2\ \text{for }\beta(g), \\
1.5+N/2\ \text{for }\gamma(g). 
\end{array}
\right.$$

\subsection{Resummed results}

Resummation procedure was  performed in two stages.
First, we determined $v_*$ as the first positive null of resummed function $\beta = \beta(v)$. Then we used this value $v_*$ in the resummed series for $\eta = \eta(v)$.
In the Table \ref{tab:vstar} we present the values of $v_*$ for $N=0$ and $N=1$ for 2-5 loop beta-function. Note that in the renormalization scheme which we use (as well as in ZM scheme), the value of  $g_*\equiv  24\pi \, v_*/(N+8)$ coincides with universal amplitude ratio (see. Appendix~\ref{appendix_gstar}), therefore it can be compared with the independent calculations that do not use the RG method (e.g., from high-temperature-series estimates). These values are in good agreement with the results obtained earlier in~\cite{Orlov2000}  using Pade-Borel resummation, but both values are higher than the estimations based on high temperature expansion~\cite{Pelissetto1998} (see also~\cite{Pelissetto2002}). 
\begin{table}[h!]
\setlength\extrarowheight{3pt}
\begin{tabular}{p{1.4cm}p{1.2cm}p{1.2cm}p{1.2cm}p{1.2cm}|p{1.6cm}p{1.6cm}}
 loops & $\quad 2$ & $\quad 3$ & $\quad 4$ & $\quad 5$ & \ \ PB & HT-exp\\
 \hline
 $N=0$ & 2.217 & 1.936 & 1.898 & 1.853 & \ 1.86(4) & 1.679(1) \\
 $N=1$ & 2.154 & 1.901 & 1.871 & 1.833 & \ 1.837(30)& 1.754(8)
\end{tabular}
\caption{Values of the fixed point $v_*$ for different loop numbers compared with 5-loop Pade-Borel resummation (PB)~\cite{Orlov2000} and estimations based on high temperature expansion (HT-exp)~\cite{Pelissetto1998}}
\label{tab:vstar}
\end{table}
These values of the fixed point were used in the resummation of the critical exponent~$\eta$. For the resummation of the six loop approximation of the exponent $\eta$ we use $v_*$ from the five loop approximation. These results in the values of the critical exponent are presented in Table~\ref{tab:eta}

\begin{table}[h]
\setlength\extrarowheight{3pt}
\begin{tabular}{p{1.4cm}p{1.2cm}p{1.2cm}p{1.2cm}p{1.2cm}p{1.4cm}}
 loops & $\quad 3$ & $\quad 4$ & $\quad 5$ & $\quad 6$ &exact \\
 \hline
 $N=0$ & 0.1022 & 0.1162 & 0.1198 & 0.1296 & $\simeq 0.208$\\
 $N=1$ & 0.1168& 0.1334 & 0.1382 & 0.1490 & 0.25 
\end{tabular}
\caption{Resummed values for critical exponent $\eta$ for different number of loops compared with the exact values $5/24$ and $1/4$ for $N=0$ and $N=1$ correspondingly.}
\label{tab:eta}
\end{table}



%% file: appendix_sd.tex
Consider an example of using of sector decomposition method in the final integral from~\eqref{sd_example}:
\be
\label{app_sd}
I_a=\int \limits_{0}^{1}\dots\int \limits_{0}^{1}du_{1}\dots du_{4} \int \limits_{0}^{1}da \left(1-a\right)
            \frac{\delta \left(\sum \limits_{i=1}^{4}u_{i} -1\right) \left(u_{1} u_{2} u_{3} u_{4} \right)^{2}} {\left(u_1 u_2 u_3 +u_1 u_2 u_4 +u_1 u_3 u_4 +a u_2 u_3 u_4 \right)^{d/2+2}}.
\ee

In section~\ref{sd_scheme} there was described the case with no stretching parameters, i.e., when $a=1$. As it was mentioned there, the decomposition in each sector ends when one of the terms of determinant becomes equal to $1$. 

In~\eqref{app_sd}, in sectors with the main variables $u_2,u_3,u_4$ (in any order), in the last term of the denominator instead of $1$ the stretching parameter $a$ appears. As there is the point where $a=0$ in the integration domain, the aim of decomposition can not be considered achieved. Therefore additional steps of decomposition are needed.

Consider as an example of such a situation sector $[u_2,u_3,u_4]$.
Performing decomposition in a similar way as in~\eqref{sd_J_2_eye}, we get: 
\be
\sHat_2 I_a = \int\limits_{0}^{1}du_1\int\limits_{0}^{1}du_3\int\limits_{0}^{1}du_4\int\limits_{0}^{1}
	da \frac{(1-a)\left(u_{1} u_{3} u_{4} \right)^{2}}
	{\left(u_1 u_3 +u_1 u_4 + u_1 u_3 u_4 +a u_3 u_4 \right)^{d/2+2}(1+u_1+u_3+u_4)^{6-3d/2}},
\ee

\be
\sHat_{23} I_a = \int\limits_{0}^{1}du_1\int\limits_{0}^{1}du_3\int\limits_{0}^{1}du_4\int\limits_{0}^{1}
	da \frac{(1-a)\left(u_{1} u_{4} \right)^{2}u_{3}^{4-d}}
	{(u_1 +u_1 u_4 + u_1 u_3 u_4 +a u_4)^{d/2+2}(1+u_1 u_3+u_3+u_3 u_4)^{6-3d/2}},
\ee

\be
\sHat_{234} I_a = \int\limits_{0}^{1}du_1\int\limits_{0}^{1}du_3\int\limits_{0}^{1}du_4\int\limits_{0}^{1}
	da \frac{(1-a)u_{1}^2 u_{3}^{4-d} u_{4}^{3-d/2}}
	{(u_1 +u_1 u_4 + u_1 u_3 u_4 + a)^{d/2+2}(1+u_1 u_3 u_4 + u_3 + u_3 u_4)^{6-3d/2}}.
\ee

In the determinant $det_{2,3,4}=(u_1 +u_1 u_4 + u_1 u_3 u_4 + a)$ instead of 1 in the final position we have stretching parameter $a$. This leads to the fact, that determinant vanishes when $u_1=0$ and $a=0$ simultaneously. To eliminate this defect, it is enough to make a decomposition step in the domain of decomposition $\{u_1,a\}$. Finally, we obtain:
\be
\sHat_a \left(\sHat_{234} I_a\right) =
\int\limits_{0}^{1}du_1\int\limits_{0}^{1}du_3\int\limits_{0}^{1}du_4\int\limits_{0}^{1}
	da \frac{a^{1-d/2}(1-a)u_{1}^2 u_{3}^{4-d} u_{4}^{3-d/2}}
	{(u_1 +u_1 u_4 + u_1 u_3 u_4 + 1)^{d/2+2}(1+a\,u_1 u_3 u_4 + u_3 + u_3 u_4)^{6-3d/2}},
\ee

\be
\sHat_1 \left(\sHat_{234} I_a\right) =
\int\limits_{0}^{1}du_1\int\limits_{0}^{1}du_3\int\limits_{0}^{1}du_4\int\limits_{0}^{1}
	da \frac{(1-a\,u_1)u_{1}^{1-d/2} u_{3}^{4-d} u_{4}^{3-d/2}}
	{(1 +u_4 + u_3 u_4 + a)^{d/2+2}(1+u_1 u_3 u_4 + u_3 + u_3 u_4)^{6-3d/2}}.
\ee
The goal of decomposition is achieved.

%% file: appendix_gstar.tex
Let us define quantity $A^R$ as
\begin{equation}\label{AR}
A^R(m,\mu,g)=\frac{-\Gamma^R_4}{\Gamma^R_2\partial_{p^2}\Gamma^R_2}\,.
\end{equation}
All renormalized Green functions here are taken at zero external momenta. At space dimension $d=2$ this quantity is dimensionless, so we can write:
\begin{equation}\label{AR1}
A^R(m,\mu,g)=A^R(t,g), \qquad t\equiv\frac{m}{\mu}.
\end{equation}
Taking into account (\ref{cond}) we can find normalization condition for $A^R(t,g)$:
\begin{equation}\label{ARnorm}
A^R(t=1,g)=g\,.
\end{equation}
Following (\ref{RGeq}) let us write renormalization group equation for $A^R(m,\mu,g)$. Due to the definition of the quantity (\ref{AR}) right hand side becomes equal to zero:
\begin{equation}\label{AReq}
 (\mu\partial_\mu+\beta \partial_g-\gamma_{m^2} m^2
\partial_{m^2})A^R(m,\mu,g)=0.  
\end{equation}
From (\ref{AR1}), (\ref{AReq}) we can derive:
\begin{equation}\label{AR1eq}
(\beta \partial_g-(1+\gamma_{m^2}/2)t\partial_t 
)A^R(t,g)=0.
\end{equation}

The last equation combined with (\ref{ARnorm}) coincide with definition of the invariant charge $\bar g(g,t)$ \cite{Vasilev2004}. If invariant charge has stable infrared fixed point $\bar g(g,t)\to g_*$ when $t \to 0$, then
\begin{equation}
A^R(t,g)\to g_*, \quad\textnormal{when}\quad  t \to 0.
\end{equation}
Nonrenormalized quantity $A$ has the same limit due to the fact of the cancellation of the renormalization constants $Z_\varphi$ in \eqref{AR}.

%% file: diagTable_5loops.tex
\begin{center}
\begingroup 
\setlength\extrarowheight{3pt}
\begin{longtable}{llllcl}
\hline\hline
  & $\gamma$ & $KR'_2$($\gamma$) & $KR'$($\gamma$) & $\ Sym(\gamma)\ $ & $3^{n+1}\times O_N$\\ 
 \hline 
1 & $ee11|ee|$ &\ 1.0(0) &\ 1.0(0) & 3/2 & $N + 8$ \\
\hline
2$^+$ & $ee11|22|ee|$ &\ 1.0(0) & -1.0(0) & 3/4 & $N^{2} + 6 N + 20$ \\
3$^*$ & $ee12|e22|e|$ &\ 0.7813024128965(10) & -0.2186975871035(10) & 3 & $5 N + 22$  \\
\hline
4$^+$ & $ee11|22|33|ee|$ &\ 1.0(0) &\ 1.0(0) & 3/8 & $N^{3} + 8 N^{2} + 24 N + 48$ \\
5$^+$ & $ee11|23|e33|e|$ &\ 0.7813024128965(10) &\ 0.2186975871035(10) & 3/2 & $3 N^{2} + 22 N + 56$ \\
6 & $ee12|ee3|333||$ &\ 0.01261399161(13) &\ 0.01261399161(13) & 1/2 & $3 N^{2} + 30 N + 48$ \\
7 & $ee12|e23|33|e|$ &\ 0.569829440(6) &\ 0.007224614(6) & 6 & $N^{2} + 20 N + 60$  \\
8$^*$ & $ee12|e33|e33||$ &\ 0.6590435620646(10) &\ 0.0964387362717(22) & 3/2 & $3 N^{2} + 22 N + 56$ \\
9 & $ee12|223|3|ee|$ &\ 0.650899895(6) &\ 0.088295069(6) & 3/4 & $3 N^{2} + 22 N + 56$ \\
10$^*$ & $e112|e3|e33|e|$ &\ 0.6590435620646(10) &\ 0.0964387362717(22) & 3/2 & $N^{2} + 20 N + 60$ \\
11 & $e123|e23|e3|e|$ &\ 0.400685635(4) &\ 0.400685635(4) & 1 & $15 N + 66$  \\
\hline
12$^+$ & $ee11|22|33|44|ee|$ &\ 1.0(0) & -1.0(0) & 3/16 & $N^{4} + 10 N^{3} + 40 N^{2} + 80 N + 112$ \\
13$^+$ & $ee11|22|34|e44|e|$ &\ 0.7813024128965(10) & -0.2186975871035(10) & 3/4 & $3 N^{3} + 24 N^{2} + 80 N + 136$ \\
14$^+$ & $ee11|23|ee4|444||$ &\ 0.01261399161(13) & -0.01261399161(13) & 1/2 & $3 N^{3} + 24 N^{2} + 96 N + 120$ \\
15$^+$ & $ee11|23|e34|44|e|$ &\ 0.569829440(6) & -0.007224614(6) & 3 & $N^{3} + 14 N^{2} + 76 N + 152$  \\
16$^+$ & $ee11|23|e44|e44||$ &\ 0.6590435620646(10) & -0.0964387362717(22) & 3/4 & $N^{3} + 18 N^{2} + 80 N + 144$ \\
17$^+$ & $ee11|23|334|4|ee|$ &\ 0.650899895(6) & -0.088295069(6) & 3/4 & $3 N^{3} + 24 N^{2} + 80 N + 136$ \\
18 & $ee12|ee3|344|44||$ &\ 0.0220353950(4) & -0.0031925883(4) & 3/4 & $N^{3} + 18 N^{2} + 96 N + 128$ \\
19 & $ee12|e23|e4|444||$ &\ 0.01299093503(27) &\ 0.00037694342(30) & 1 & $15 N^{2} + 96 N + 132$  \\
20 & $ee12|e23|34|44|e|$ &\ 0.407546072(4) &\ 0.001360971(12) & 6 & $7 N^{2} + 72 N + 164$  \\
21 & $ee12|e23|44|e44||$ &\ 0.459171682(5) &\ 0.004376479(12) & 3 & $N^{3} + 14 N^{2} + 76 N + 152$  \\
22$^*$ & $ee12|e33|e44|44||$ &\ 0.5761437790903(10) & -0.057079668414(4) & 3/4 & $3 N^{3} + 24 N^{2} + 80 N + 136$ \\
23 & $ee12|e33|344|4|e|$ &\ 0.488684862(5) & -0.006714361(7) & 3 & $11 N^{2} + 76 N + 156$  \\
24 & $ee12|e33|444|e4||$ &\ 0.0055553718(4) & -0.0070586198(4) & 1 & $15 N^{2} + 96 N + 132$  \\
25 & $ee12|e34|e34|44||$ &\ 0.409355762(4) &\ 0.003755035(8) & 3/2 & $11 N^{2} + 76 N + 156$ \\
26 & $ee12|e34|334|4|e|$ &\ 0.383792310(4) & -0.022392791(12) & 6 & $7 N^{2} + 72 N + 164$  \\
27 & $ee12|223|4|e44|e|$ &\ 0.542459696(5) & -0.034009983(8) & 3/2 & $11 N^{2} + 76 N + 156$ \\
28 & $ee12|233|34|4|ee|$ &\ 0.440797049(4) & -0.005854486(14) & 3/2 & $N^{3} + 14 N^{2} + 76 N + 152$ \\
29 & $ee12|233|44|e4|e|$ &\ 0.459171682(5) &\ 0.004376480(12) & 3/2 & $N^{3} + 10 N^{2} + 72 N + 160$ \\
30 & $ee12|234|34|e4|e|$ &\ 0.3126232707(31) & -0.088062364(5) & 3 & $5 N^{2} + 62 N + 176$  \\
31 & $ee12|334|334||ee|$ &\ 0.487129469(5) & -0.007547793(14) & 3/8 & $N^{3} + 18 N^{2} + 80 N + 144$ \\
32 & $ee12|334|344|e|e|$ &\ 0.449316700(4) & -0.005478503(12) & 3/2 & $7 N^{2} + 72 N + 164$ \\
33$^+$ & $e112|e2|34|e44|e|$ &\ 0.6104334603979(16) & -0.0478286346049(4) & 3/4 & $11 N^{2} + 76 N + 156$ \\
34 & $e112|e3|e34|44|e|$ &\ 0.488684862(5) & -0.006714362(7) & 6 & $7 N^{2} + 72 N + 164$  \\
35$^*$ & $e112|e3|e44|e44||$ &\ 0.5761437790903(10) & -0.057079668414(4) & 3/2 & $N^{3} + 14 N^{2} + 76 N + 152$ \\
36 & $e112|34|e34|e4|e|$ &\ 0.320802341(12) & -0.079883294(12) & 6 & $5 N^{2} + 62 N + 176$  \\
37 & $e123|e24|34|e4|e|$ &\ 0.262808539(9) &\ 0.262808539(9) & 3 & $2 N^{2} + 55 N + 186$  \\
\hline
38$^+$  & $ee11|22|33|44|55|ee|$ &\ 1.0(0) &\ 1.0(0) & 3/32 & $N^{5} + 12 N^{4} + 60 N^{3} + 160 N^{2}+$\\ & & & & & $\qquad\qquad\qquad\qquad + 240 N + 256$ \\
39$^+$  & $ee11|22|33|45|e55|e|$ &\ 0.7813024128965(10) &\ 0.2186975871035(10) & 3/8 & $3 N^{4} + 30 N^{3} + 120 N^{2} + 256 N + 320$ \\
40$^+$  & $ee11|22|34|ee5|555||$ &\ 0.01261399161(13) &\ 0.01261399161(13) & 1/4 & $3 N^{4} + 30 N^{3} + 120 N^{2} + 288 N + 288$ \\
41$^+$  & $ee11|22|34|e45|55|e|$ &\ 0.569829440(6) &\ 0.007224614(6) & 3/2 & $N^{4} + 16 N^{3} + 88 N^{2} + 256 N + 368$ \\
42$^+$  & $ee11|22|34|e55|e55||$ &\ 0.6590435620646(10) &\ 0.0964387362717(22) & 3/8 & $N^{4} + 16 N^{3} + 96 N^{2} + 264 N + 352$ \\
43$^+$  & $ee11|22|34|445|5|ee|$ &\ 0.650899895(6) &\ 0.088295069(6) & 3/8 & $3 N^{4} + 30 N^{3} + 120 N^{2} + 256 N + 320$ \\
44$^+$  & $ee11|23|ee4|455|55||$ &\ 0.0220353950(4) &\ 0.0031925883(4) & 3/4 & $N^{4} + 16 N^{3} + 96 N^{2} + 296 N + 320$ \\
45$^+$  & $ee11|23|e34|e5|555||$ &\ 0.01299093503(27) & -0.00037694342(30) & 1/2 & $9 N^{3} + 84 N^{2} + 300 N + 336$ \\
46$^+$  & $ee11|23|e34|45|55|e|$ &\ 0.407546072(4) & -0.001360971(12) & 3 & $5 N^{3} + 56 N^{2} + 252 N + 416$  \\
47$^+$  & $ee11|23|e34|55|e55||$ &\ 0.459171682(5) & -0.004376479(12) & 3/2 & $N^{4} + 12 N^{3} + 76 N^{2} + 256 N + 384$ \\
48$^+$  & $ee11|23|e44|e55|55||$ &\ 0.5761437790903(10) &\ 0.057079668414(4) & 3/8 & $N^{4} + 16 N^{3} + 96 N^{2} + 264 N + 352$ \\
49$^+$  & $ee11|23|e44|455|5|e|$ &\ 0.488684862(5) &\ 0.006714361(7) & 3/2 & $5 N^{3} + 64 N^{2} + 260 N + 400$ \\
50$^+$  & $ee11|23|e44|555|e5||$ &\ 0.0055553718(4) &\ 0.0070586198(4) & 1/2 & $9 N^{3} + 84 N^{2} + 300 N + 336$ \\
51$^+$  & $ee11|23|e45|e45|55||$ &\ 0.409355762(4) & -0.003755035(8) & 3/4 & $5 N^{3} + 64 N^{2} + 260 N + 400$ \\
52$^+$  & $ee11|23|e45|445|5|e|$ &\ 0.383792310(4) &\ 0.022392791(12) & 3 & $5 N^{3} + 56 N^{2} + 252 N + 416$  \\
53$^+$  & $ee11|23|334|4|55|ee|$ &\ 0.650899895(6) &\ 0.088295069(6) & 3/16 & $3 N^{4} + 30 N^{3} + 120 N^{2} + 256 N + 320$ \\
54$^+$  & $ee11|23|334|5|e55|e|$ &\ 0.542459696(5) &\ 0.034009983(8) & 3/4 & $9 N^{3} + 76 N^{2} + 260 N + 384$ \\
55$^+$  & $ee11|23|344|45|5|ee|$ &\ 0.440797049(4) &\ 0.005854486(14) & 3/2 & $N^{4} + 16 N^{3} + 88 N^{2} + 256 N + 368$ \\
56$^+$  & $ee11|23|344|55|e5|e|$ &\ 0.459171682(5) & -0.004376480(12) & 3/4 & $N^{4} + 12 N^{3} + 68 N^{2} + 248 N + 400$ \\
57$^+$  & $ee11|23|345|45|e5|e|$ &\ 0.3126232707(31) &\ 0.088062364(5) & 3/2 & $5 N^{3} + 52 N^{2} + 232 N + 440$ \\
58$^+$  & $ee11|23|444|455||ee|$ &\ 0.01261399161(13) &\ 0.01261399161(13) & 1/8 & $3 N^{4} + 30 N^{3} + 120 N^{2} + 288 N + 288$ \\
59$^+$  & $ee11|23|445|445||ee|$ &\ 0.487129469(5) &\ 0.007547793(14) & 3/8 & $N^{4} + 16 N^{3} + 96 N^{2} + 264 N + 352$ \\
60$^+$  & $ee11|23|445|455|e|e|$ &\ 0.449316700(4) &\ 0.005478503(12) & 3/4 & $5 N^{3} + 56 N^{2} + 252 N + 416$ \\
61  & $ee12|ee3|334|5|555||$ & -6.08874(13)e-05 & -6.08874(13)e-05 & 1/4 & $9 N^{3} + 108 N^{2} + 324 N + 288$ \\
62$^*$  & $ee12|ee3|344|55|55||$ &\ 0.0286150341(10) &\ 0.0003508240(15) & 3/8 & $N^{4} + 16 N^{3} + 96 N^{2} + 296 N + 320$ \\
63  & $ee12|ee3|345|45|55||$ &\ 0.0159249(4) & -0.0005932(4) & 3/4 & $5 N^{3} + 72 N^{2} + 300 N + 352$ \\
64 & $ee12|ee3|444|555|5||$ &\ 0.000571523(35) &\ 0.000571523(35) & 1/12 & $9 N^{3} + 108 N^{2} + 324 N + 288$ \\
65 & $ee12|ee3|445|455|5||$ &\ 0.02809324(8) &\ 0.00215374(8) & 3/4 & $5 N^{3} + 72 N^{2} + 300 N + 352$ \\
66 & $ee12|e23|e4|455|55||$ &\ 0.02254724(8) & -0.00024204(8) & 3/2 & $5 N^{3} + 72 N^{2} + 300 N + 352$ \\
67 & $ee12|e23|34|e5|555||$ &\ 0.01015606(9) & -7.622(9)e-05 & 2 & $3 N^{3} + 66 N^{2} + 300 N + 360$  \\
68 & $ee12|e23|34|45|55|e|$ &\ 0.2899253(8) & -6.0(8)e-06 & 6 & $N^{3} + 36 N^{2} + 244 N + 448$  \\
69 & $ee12|e23|34|55|e55||$ &\ 0.3244556(19) & -0.0005314(19) & 3 & $5 N^{3} + 56 N^{2} + 252 N + 416$  \\
70$^*$ & $ee12|e23|44|e55|55||$ &\ 0.3880975799240(10) & -0.004151891(22) & 3/2 & $N^{4} + 16 N^{3} + 88 N^{2} + 256 N + 368$ \\
71 & $ee12|e23|44|455|5|e|$ &\ 0.3357750(9) & -0.0006476(9) & 3 & $3 N^{3} + 50 N^{2} + 252 N + 424$  \\
72 & $ee12|e23|44|555|e5||$ &\ 0.0054562(6) & -0.0004761(6) & 1 & $3 N^{3} + 66 N^{2} + 300 N + 360$  \\
73 & $ee12|e23|45|e45|55||$ &\ 0.2816184(11) & -0.0010528(11) & 3 & $3 N^{3} + 50 N^{2} + 252 N + 424$  \\
74 & $ee12|e23|45|445|5|e|$ &\ 0.2683150(12) &\ 0.0021375(12) & 6 & $N^{3} + 36 N^{2} + 244 N + 448$  \\
75 & $ee12|e33|e34|5|555||$ &\ 0.01024426(7) &\ 1.198(7)e-05 & 1 & $9 N^{3} + 84 N^{2} + 300 N + 336$  \\
76$^*$ & $ee12|e33|e44|55|55||$ &\ 0.5146991310775(10) &\ 0.039175735518(8) & 3/8 & $3 N^{4} + 30 N^{3} + 120 N^{2} + 256 N + 320$ \\
77 & $ee12|e33|e45|45|55||$ &\ 0.3609855(10) &\ 0.0003640(10) & 3/2 & $9 N^{3} + 76 N^{2} + 260 N + 384$ \\
78 & $ee12|e33|344|5|55|e|$ &\ 0.3973679(7) & -0.0006826(7) & 3/2 & $5 N^{3} + 64 N^{2} + 260 N + 400$ \\
79 & $ee12|e33|345|4|55|e|$ &\ 0.3509568(8) & -0.0006024(8) & 3 & $3 N^{3} + 50 N^{2} + 252 N + 424$  \\
80 & $ee12|e33|444|55|5|e|$ &\ 0.00321810(4) &\ 0.00472135(4) & 1/2 & $9 N^{3} + 84 N^{2} + 300 N + 336$ \\
81 & $ee12|e33|445|e5|55||$ &\ 0.0102191(4) &\ 0.0023010(4) & 3/2 & $5 N^{3} + 72 N^{2} + 300 N + 352$ \\
82 & $ee12|e33|445|45|5|e|$ &\ 0.33492398(32) &\ 0.00711854(32) & 3 & $3 N^{3} + 50 N^{2} + 252 N + 424$  \\
83 & $ee12|e33|445|55|e5||$ &\ 0.4312636(4) &\ 0.0054552(4) & 3/2 & $9 N^{3} + 76 N^{2} + 260 N + 384$ \\
84 & $ee12|e34|e34|55|55||$ &\ 0.2977295(7) & -0.0050863(7) & 3/4 & $5 N^{3} + 64 N^{2} + 260 N + 400$ \\
85 & $ee12|e34|e35|45|55||$ &\ 0.26784902(21) & -0.00281024(21) & 3 & $3 N^{3} + 50 N^{2} + 252 N + 424$  \\
86 & $ee12|e34|334|5|55|e|$ &\ 0.2974294(4) &\ 0.0060512(4) & 3 & $N^{3} + 36 N^{2} + 244 N + 448$  \\
87 & $ee12|e34|335|e|555||$ &\ 0.00620152(7) &\ 0.00026920(7) & 1 & $3 N^{3} + 66 N^{2} + 300 N + 360$  \\
88 & $ee12|e34|335|4|55|e|$ &\ 0.3012655(5) &\ 3.22(5)e-05 & 3 & $3 N^{3} + 42 N^{2} + 244 N + 440$  \\
89 & $ee12|e34|335|5|e55||$ &\ 0.3155247(5) &\ 0.0028559(5) & 3 & $3 N^{3} + 50 N^{2} + 252 N + 424$  \\
90 & $ee12|e34|345|e5|55||$ &\ 0.2510146(5) &\ 0.0085908(5) & 6 & $N^{3} + 36 N^{2} + 244 N + 448$  \\
91 & $ee12|e34|345|45|5|e|$ &\ 0.2087328(5) & -0.0162615(5) & 6 & $25 N^{2} + 220 N + 484$  \\
92 & $ee12|e34|355|e4|55||$ &\ 0.2913870(5) &\ 0.0139076(5) & 3 & $5 N^{3} + 56 N^{2} + 252 N + 416$  \\
93 & $ee12|e34|355|44|5|e|$ &\ 0.3262784(7) & -0.0003051(7) & 3 & $N^{3} + 44 N^{2} + 252 N + 432$  \\
94 & $ee12|e34|355|45|e5||$ &\ 0.2590697(11) & -0.0071078(11) & 6 & $N^{3} + 36 N^{2} + 244 N + 448$  \\
95 & $ee12|e34|555|e44|5||$ &\ 0.003437955(34) &\ 0.000641232(34) & 1 & $3 N^{3} + 66 N^{2} + 300 N + 360$  \\
96$^+$  & $ee12|223|3|45|e55|e|$ &\ 0.508549659(5) &\ 0.0193099186(14) & 3/4 & $9 N^{3} + 76 N^{2} + 260 N + 384$ \\
97 & $ee12|223|4|ee5|555||$ &\ 0.005098474(4) &\ 0.002301752(4) & 1/2 & $9 N^{3} + 84 N^{2} + 300 N + 336$ \\
98 & $ee12|223|4|e45|55|e|$ &\ 0.400895605(13) &\ 0.001855900(15) & 3 & $3 N^{3} + 50 N^{2} + 252 N + 424$  \\
99 & $ee12|223|4|e55|e55||$ &\ 0.470555875(35) &\ 0.01826823(4) & 3/4 & $5 N^{3} + 64 N^{2} + 260 N + 400$ \\
100 & $ee12|223|4|445|5|ee|$ &\ 0.447721788(4) &\ 0.012637161(12) & 3/8 & $9 N^{3} + 76 N^{2} + 260 N + 384$ \\
101 & $ee12|233|34|5|e55|e|$ &\ 0.37438904(5) &\ 0.00249793(5) & 3 & $3 N^{3} + 50 N^{2} + 252 N + 424$  \\
102$^*$ & $ee12|233|44|e5|55|e|$ &\ 0.3880975799240(10) & -0.004151892(20) & 3/2 & $N^{4} + 12 N^{3} + 68 N^{2} + 248 N + 400$ \\
103 & $ee12|233|44|45|5|ee|$ &\ 0.33706660(6) &\ 0.00118070(7) & 3/8 & $N^{4} + 12 N^{3} + 68 N^{2} + 248 N + 400$ \\
104 & $ee12|233|45|e4|55|e|$ &\ 0.3357750(17) & -0.0006476(17) & 3 & $3 N^{3} + 42 N^{2} + 244 N + 440$  \\
105 & $ee12|233|45|44|5|ee|$ &\ 0.337066617(21) &\ 0.001180718(34) & 3/4 & $N^{4} + 12 N^{3} + 76 N^{2} + 256 N + 384$ \\
106 & $ee12|233|45|45|e5|e|$ &\ 0.2374371(22) &\ 0.0046971(22) & 3 & $3 N^{3} + 38 N^{2} + 224 N + 464$  \\
107 & $ee12|234|34|e5|55|e|$ &\ 0.2481407(12) &\ 0.0154007(12) & 3 & $3 N^{3} + 38 N^{2} + 224 N + 464$  \\
108 & $ee12|234|34|45|5|ee|$ &\ 0.246028090(23) &\ 0.021467183(24) & 3/4 & $5 N^{3} + 52 N^{2} + 232 N + 440$ \\
109 & $ee12|234|35|ee|555||$ &\ 0.0129040740(28) & -0.0004638045(28) & 1/4 & $9 N^{3} + 84 N^{2} + 300 N + 336$ \\
110 & $ee12|234|35|e4|55|e|$ &\ 0.2451429(15) &\ 0.0124029(15) & 6 & $N^{3} + 32 N^{2} + 224 N + 472$  \\
111 & $ee12|234|35|44|5|ee|$ &\ 0.291105769(14) &\ 0.006466162(23) & 3 & $5 N^{3} + 56 N^{2} + 252 N + 416$  \\
112 & $ee12|234|35|45|e5|e|$ &\ 0.2016853(14) & -0.0611232(14) & 6 & $N^{3} + 26 N^{2} + 210 N + 492$  \\
113$^+$  & $ee12|333|345||e55|e|$ &\ 0.00985534208(10) &\ 0.002758649529(28) & 1/2 & $9 N^{3} + 84 N^{2} + 300 N + 336$ \\
114 & $ee12|333|444|5|5|ee|$ &\ 0.000571524(11) &\ 0.000571524(11) & 1/24 & $9 N^{3} + 108 N^{2} + 324 N + 288$ \\
115 & $ee12|333|445|5|e5|e|$ &\ 0.00343796(12) &\ 0.00064124(12) & 1 & $3 N^{3} + 66 N^{2} + 300 N + 360$  \\
116 & $ee12|334|335||e55|e|$ &\ 0.418429182(23) &\ 0.007653111(29) & 3/4 & $5 N^{3} + 64 N^{2} + 260 N + 400$ \\
117 & $ee12|334|344|5|5|ee|$ &\ 0.323263628(11) &\ 0.007087693(29) & 3/8 & $5 N^{3} + 56 N^{2} + 252 N + 416$ \\
118 & $ee12|334|345|e|55|e|$ &\ 0.3265574(7) & -1.02(7)e-05 & 3 & $N^{3} + 36 N^{2} + 244 N + 448$  \\
119 & $ee12|334|345|5|e5|e|$ &\ 0.2299142(15) & -0.0028258(15) & 6 & $N^{3} + 32 N^{2} + 224 N + 472$  \\
120 & $ee12|334|355|e|e55||$ &\ 0.3744262(14) &\ 0.0018867(14) & 3/2 & $5 N^{3} + 56 N^{2} + 252 N + 416$ \\
121 & $ee12|334|355|4|e5|e|$ &\ 0.3155246(12) &\ 0.0028558(12) & 3 & $3 N^{3} + 42 N^{2} + 244 N + 440$  \\
122 & $ee12|334|355|5|ee5||$ &\ 0.339430552(13) & -0.000514701(22) & 3/4 & $5 N^{3} + 64 N^{2} + 260 N + 400$ \\
123 & $ee12|334|455|e4|5|e|$ &\ 0.3638837(11) &\ 0.0008889(11) & 3/2 & $N^{3} + 44 N^{2} + 252 N + 432$ \\
124 & $ee12|334|455|e5|e5||$ &\ 0.3119080(17) &\ 0.0090942(17) & 3 & $N^{3} + 36 N^{2} + 244 N + 448$  \\
125 & $ee12|334|455|55|ee||$ &\ 0.390252839(14) & -0.004276548(28) & 3/16 & $N^{4} + 16 N^{3} + 96 N^{2} + 264 N + 352$ \\
126 & $ee12|345|345|ee|55||$ &\ 0.292660934(30) & -0.005573202(33) & 3/8 & $5 N^{3} + 64 N^{2} + 260 N + 400$ \\
127 & $ee12|345|345|e4|5|e|$ &\ 0.1919035(11) & -0.0709050(11) & 3 & $19 N^{2} + 206 N + 504$  \\
128$^+$ & $e112|e2|33|45|e55|e|$ &\ 0.6104334603979(16) &\ 0.0478286346049(4) & 3/8 & $9 N^{3} + 76 N^{2} + 260 N + 384$ \\
129$^+$  & $e112|e2|34|e45|55|e|$ &\ 0.445209116(4) &\ 0.0015800056(12) & 3 & $3 N^{3} + 50 N^{2} + 252 N + 424$  \\
130$^+$  & $e112|e2|34|e55|e55||$ &\ 0.5149123252450(10) &\ 0.0210909189259(6) & 3/4 & $5 N^{3} + 64 N^{2} + 260 N + 400$ \\
131 & $e112|e3|e34|e5|555||$ &\ 0.010244267(20) &\ 1.1982(20)e-05 & 1 & $3 N^{3} + 66 N^{2} + 300 N + 360$  \\
132 & $e112|e3|e34|45|55|e|$ &\ 0.3509570(10) & -0.0006022(10) & 6 & $N^{3} + 36 N^{2} + 244 N + 448$  \\
133 & $e112|e3|e34|55|e55||$ &\ 0.3973679(24) & -0.0006826(24) & 3 & $5 N^{3} + 56 N^{2} + 252 N + 416$  \\
134$^*$ & $e112|e3|e44|e55|55||$ &\ 0.5146991310775(10) &\ 0.039175735518(7) & 3/4 & $N^{4} + 16 N^{3} + 88 N^{2} + 256 N + 368$ \\
135 & $e112|e3|e44|455|5|e|$ &\ 0.4312637(9) &\ 0.0054553(9) & 3 & $3 N^{3} + 50 N^{2} + 252 N + 424$  \\
136 & $e112|e3|e44|555|e5||$ &\ 0.0032181(4) &\ 0.0047214(4) & 1/2 & $3 N^{3} + 66 N^{2} + 300 N + 360$ \\
137 & $e112|e3|e45|e45|55||$ &\ 0.3609855(12) &\ 0.0003640(12) & 3/2 & $3 N^{3} + 50 N^{2} + 252 N + 424$ \\
138 & $e112|e3|e45|445|5|e|$ &\ 0.3349240(12) &\ 0.0071185(12) & 6 & $N^{3} + 36 N^{2} + 244 N + 448$  \\
139 & $e112|e3|334|5|e55|e|$ &\ 0.45340530(5) &\ 0.01373900(5) & 3/4 & $N^{3} + 44 N^{2} + 252 N + 432$ \\
140 & $e112|e3|344|55|e5|e|$ &\ 0.397368(4) & -0.000682(4) & 3/2 & $3 N^{3} + 42 N^{2} + 244 N + 440$ \\
141 & $e112|e3|345|45|e5|e|$ &\ 0.2651800(20) &\ 0.0401857(20) & 3 & $25 N^{2} + 220 N + 484$  \\
142 & $e112|e3|445|455|e|e|$ &\ 0.3911997(16) &\ 0.0030043(16) & 3/2 & $N^{3} + 36 N^{2} + 244 N + 448$ \\
143 & $e112|23|e4|e45|55|e|$ &\ 0.3638837(8) &\ 0.0008889(8) & 3 & $N^{3} + 36 N^{2} + 244 N + 448$  \\
144 & $e112|23|e4|e55|e55||$ &\ 0.4312635(11) &\ 0.0054550(11) & 3 & $5 N^{3} + 56 N^{2} + 252 N + 416$  \\
145 & $e112|23|e4|455|e5|e|$ &\ 0.3638837(12) &\ 0.0008889(12) & 3 & $N^{3} + 36 N^{2} + 244 N + 448$  \\
146 & $e112|23|45|e45|e5|e|$ &\ 0.2340570(31) &\ 0.0008836(31) & 6 & $25 N^{2} + 220 N + 484$  \\
147$^*$ & $e112|33|e44|e5|55|e|$ &\ 0.5146991310775(10) &\ 0.039175735518(7) & 3/8 & $N^{4} + 12 N^{3} + 76 N^{2} + 256 N + 384$ \\
148 & $e112|33|e45|45|e5|e|$ &\ 0.269623(5) &\ 0.028704(5) & 3 & $5 N^{3} + 52 N^{2} + 232 N + 440$  \\
149 & $e112|34|e34|e5|55|e|$ &\ 0.259126(6) &\ 0.018207(6) & 3 & $N^{3} + 32 N^{2} + 224 N + 472$  \\
150 & $e112|34|e35|e4|55|e|$ &\ 0.262290(6) &\ 0.021371(6) & 3 & $3 N^{3} + 38 N^{2} + 224 N + 464$  \\
151 & $e112|34|e35|e5|e55||$ &\ 0.262290(4) &\ 0.021371(4) & 6 & $N^{3} + 32 N^{2} + 224 N + 472$  \\
152 & $e112|34|e35|45|e5|e|$ &\ 0.2140141(21) & -0.0487944(21) & 12 & $N^{3} + 26 N^{2} + 210 N + 492$  \\
153 & $e112|34|e55|e45|e5||$ &\ 0.269614(8) &\ 0.028695(8) & 3/2 & $3 N^{3} + 38 N^{2} + 224 N + 464$ \\
154 & $e112|34|345|e5|e5|e|$ &\ 0.2073468(28) & -0.0554617(28) & 6 & $19 N^{2} + 206 N + 504$  \\
155 & $e123|e23|e4|e5|555||$ &\ 0.0039545(13) &\ 0.0039545(13) & 1 & $45 N^{2} + 288 N + 396$  \\
156 & $e123|e23|45|45|e5|e|$ &\ 0.1674887(33) &\ 0.1674887(33) & 3/2 & $15 N^{2} + 186 N + 528$ \\
157 & $e123|e24|e5|e45|55||$ &\ 0.2160650(30) & -0.0171084(30) & 6 & $25 N^{2} + 220 N + 484$  \\
158 & $e123|e24|35|45|e5|e|$ &\ 0.1674053(16) &\ 0.1674053(16) & 12 & $14 N^{2} + 189 N + 526$  \\
159 & $e123|e24|55|e45|e5||$ &\ 0.197780(4) & -0.065029(4) & 3 & $N^{3} + 26 N^{2} + 210 N + 492$  \\
160 & $e123|e45|e45|e45|5||$ &\ 0.1612824(14) &\ 0.1612824(14) & 2 & $15 N^{2} + 186 N + 528$  \\
161 & $e123|e45|e45|445||e|$ &\ 0.1884825(35) & -0.0743260(35) & 3/2 & $19 N^{2} + 206 N + 504$ \\
\hline \hline 
\caption{The values of the individual vertex diagrams up to 5 loops}
\label{table_III}
\end{longtable}

\begin{longtable}{llllcl}
 \hline \hline 
  & $\gamma$ & $KR'_2$($-\partial_{p^2}\gamma$) & $KR'$($-\partial_{p^2}\gamma$) & $\ Sym(\gamma)\ $ & $3^n\times O_N/(N+2)$\\ 
 \hline 
1$^*$ & $e111|e|$ &\ 0.1146357462298(10) &\ 0.1146357462298(10) & 1/6 & $3$ \\
\hline
2$^*$ & $e112|22|e|$ &\ 0.1573984097711(10) & -0.0718730826886(22) & 1/4 & $N + 8$ \\
\hline
3 & $e112|e3|333||$ &\ 1.426675544(10)e-05 &\ 1.426675544(10)e-05 & 1/12 & $9 N + 18$ \\
4 & $e112|23|33|e|$ &\ 0.1725706567(17) &\ 0.0225507055(17) & 1/4 & $5 N + 22$ \\
5$^*$ & $e112|33|e33||$ &\ 0.1764000202033(10) &\ 0.048112029580(4) & 1/8 & $N^{2} + 6 N + 20$ \\
6 & $e123|e23|33||$ &\ 0.1080374073(11) &\ 0.0007801197(11) & 1/4 & $5 N + 22$ \\
\hline
7$^*$ & $e112|e3|344|44||$ &\ 0.00015301003(10) &\ 0.00012447652(18) & 1/8 & $3 N^{2} + 30 N + 48$ \\
8$^*$ & $e112|23|e4|444||$ &\ 0.0013753743(10) & -8.49068(10)e-05 & 1/6 & $3 N^{2} + 30 N + 48$ \\
9 & $e112|23|34|44|e|$ &\ 0.16585929(12) & -0.00483583(12) & 1/4 & $N^{2} + 20 N + 60$ \\
10 & $e112|23|44|e44||$ &\ 0.17877610(10) & -0.01197955(10) & 1/4 & $3 N^{2} + 22 N + 56$ \\
11$^*$ & $e112|33|e44|44||$ &\ 0.1848089769931(10) & -0.034943630113(8) & 1/16 & $N^{3} + 8 N^{2} + 24 N + 48$ \\
12$^*$ & $e112|33|444|e4||$ & -0.0005104144215(10) & -0.0005389479324(10) & 1/24 & $3 N^{2} + 30 N + 48$ \\
13 & $e112|34|e34|44||$ &\ 0.11737787(19) & -0.00107873(19) & 1/4 & $3 N^{2} + 22 N + 56$ \\
14 & $e112|34|334|4|e|$ &\ 0.12367221(16) & -0.00206887(16) & 1/2 & $N^{2} + 20 N + 60$ \\
15 & $e123|e23|44|44||$ &\ 0.0829725(5) &\ 0.0021854(5) & 1/8 & $3 N^{2} + 22 N + 56$ \\
16 & $e123|e24|34|44||$ &\ 0.07556620(29) &\ 0.00029504(29) & 1/2 & $N^{2} + 20 N + 60$ \\
17 & $e123|234|34|4|e|$ &\ 0.0738347(4) & -0.0180311(4) & 1/6 & $15 N + 66$ \\
\hline \hline 
\caption{The values of the individual self energy diagrams up to 5 loops}
\label{table_IV}
\end{longtable}
\endgroup
\end{center}

%% file: diagTable.tex
\begin{center}
\begingroup
\setlength\extrarowheight{3pt}
 \begin{longtable}{l c l l c l }
 \hline\hline
  & $\gamma$ & $KR'_2(-\partial_{p^2}\gamma$) & $KR'(-\partial_{p^2}\gamma)$ & $\ Sym(\gamma)\ $ & $3^6\times O_N/(N+2)$\\ 
 \hline 
1$^*$ & $e112|e3|334|5|555||$ & -6.6706(10)e-07 & -6.6706(10)e-07 & 1/24 & $27 N^{2} + 108 N + 108$ \\
2$^*$ & $e112|e3|344|55|55||$ &\ 0.00033051162(10) & -8.57182(4)e-05 & 1/16 & $3 N^{3} + 24 N^{2} + 96 N + 120$ \\
3 & $e112|e3|345|45|55||$ &\ 0.0001318(7) & -1.50(7)e-05 & 1/8 & $15 N^{2} + 96 N + 132$ \\
4$^*$ & $e112|e3|444|555|5||$ & -2.29191394(10)e-05 & -2.29191394(10)e-05 & 1/72 & $27 N^{2} + 108 N + 108$ \\
5 & $e112|e3|445|455|5||$ &\ 0.0003213(8) &\ 3.58(8)e-05 & 1/8 & $15 N^{2} + 96 N + 132$ \\
6$^*$ & $e112|23|e4|455|55||$ &\ 0.0024441251(10) & -6.51153(22)e-05 & 1/4 & $N^{3} + 18 N^{2} + 96 N + 128$ \\
7 & $e112|23|34|e5|555||$ &\ 0.0019430(13) & -1.19(13)e-05 & 1/6 & $15 N^{2} + 96 N + 132$ \\
8 & $e112|23|34|45|55|e|$ &\ 0.1489529(13) &\ 0.0009942(13) & 1/4 & $7 N^{2} + 72 N + 164$ \\
9 & $e112|23|34|55|e55||$ &\ 0.1615400(13) &\ 0.0020559(13) & 1/4 & $N^{3} + 14 N^{2} + 76 N + 152$ \\
10 & $e112|23|44|e55|55||$ &\ 0.1799854(20) &\ 0.0078676(20) & 1/8 & $3 N^{3} + 24 N^{2} + 80 N + 136$ \\
11 & $e112|23|44|455|5|e|$ &\ 0.1666787(19) &\ 0.0023645(19) & 1/8 & $11 N^{2} + 76 N + 156$ \\
12 & $e112|23|44|555|e5||$ &\ 0.0002923(15) &\ 0.0002539(15) & 1/12 & $15 N^{2} + 96 N + 132$ \\
13 & $e112|23|45|e45|55||$ &\ 0.1170513(9) & -0.0001019(9) & 1/4 & $11 N^{2} + 76 N + 156$ \\
14 & $e112|23|45|445|5|e|$ &\ 0.1170189(7) &\ 0.0010917(7) & 1/2 & $7 N^{2} + 72 N + 164$ \\
15 & $e112|33|e34|5|555||$ &\ 0.0019429(10) &\ 0.0001131(10) & 1/12 & $3 N^{3} + 24 N^{2} + 96 N + 120$ \\
16$^*$ & $e112|33|e44|55|55||$ &\ 0.1877930469979(10) &\ 0.026942997504(16) & 1/32 & $N^{4} + 10 N^{3} + 40 N^{2} + 80 N + 112$ \\
17 & $e112|33|e45|45|55||$ &\ 0.1213661(13) &\ 0.0011935(13) & 1/8 & $3 N^{3} + 24 N^{2} + 80 N + 136$ \\
18 & $e112|33|344|5|55|e|$ &\ 0.1789059(28) &\ 0.0057890(28) & 1/16 & $N^{3} + 18 N^{2} + 80 N + 144$ \\
19$^*$ & $e112|33|444|55|5|e|$ & -0.0008109801015(10) &\ 0.0007630634294(32) & 1/48 & $3 N^{3} + 24 N^{2} + 96 N + 120$ \\
20$^*$ & $e112|33|445|e5|55||$ & -0.0008074901(10) & -3.562(5)e-05 & 1/16 & $N^{3} + 18 N^{2} + 96 N + 128$ \\
21 & $e112|33|445|45|5|e|$ &\ 0.1302831(10) &\ 0.0018091(11) & 1/4 & $N^{3} + 14 N^{2} + 76 N + 152$ \\
22$^*$ & $e112|34|e33|5|555||$ &\ 0.0019427996(10) &\ 0.0001129242(22) & 1/24 & $3 N^{3} + 24 N^{2} + 96 N + 120$ \\
23 & $e112|34|e34|55|55||$ &\ 0.0888124(11) & -0.0012220(13) & 1/8 & $N^{3} + 18 N^{2} + 80 N + 144$ \\
24 & $e112|34|e35|45|55||$ &\ 0.0805963(5) & -0.0001015(6) & 1/2 & $N^{3} + 14 N^{2} + 76 N + 152$ \\
25 & $e112|34|e55|445|5||$ &\ 0.1213654(18) &\ 0.0011928(18) & 1/16 & $3 N^{3} + 24 N^{2} + 80 N + 136$ \\
26 & $e112|34|334|5|55|e|$ &\ 0.1264413(14) &\ 0.0020260(14) & 1/8 & $7 N^{2} + 72 N + 164$ \\
27 & $e112|34|335|e|555||$ &\ 0.0013090(17) &\ 0.0001909(17) & 1/24 & $15 N^{2} + 96 N + 132$ \\
28 & $e112|34|335|4|55|e|$ &\ 0.1264754(14) & -0.0001994(14) & 1/8 & $N^{3} + 10 N^{2} + 72 N + 160$ \\
29 & $e112|34|335|5|e55||$ &\ 0.1265286(10) &\ 0.0008226(10) & 1/4 & $11 N^{2} + 76 N + 156$ \\
30 & $e112|34|345|e5|55||$ &\ 0.0852574(5) &\ 0.0004023(6) & 1/2 & $7 N^{2} + 72 N + 164$ \\
31 & $e112|34|345|45|5|e|$ &\ 0.0770220(4) &\ 0.0053055(6) & 1/2 & $5 N^{2} + 62 N + 176$ \\
32 & $e112|34|355|e4|55||$ &\ 0.0971543(7) &\ 2.1(1.0)e-06 & 1/4 & $N^{3} + 14 N^{2} + 76 N + 152$ \\
33 & $e112|34|355|45|e5||$ &\ 0.0883917(5) &\ 0.0008136(6) & 1/2 & $7 N^{2} + 72 N + 164$ \\
34 & $e123|e23|34|5|555||$ &\ 0.0020532(6) & -1.86(6)e-05 & 1/12 & $15 N^{2} + 96 N + 132$ \\
35 & $e123|e23|44|55|55||$ &\ 0.0674634(11) & -0.0016534(19) & 1/16 & $3 N^{3} + 24 N^{2} + 80 N + 136$ \\
36 & $e123|e23|45|45|55||$ &\ 0.0501658(6) & -0.0003100(8) & 1/8 & $11 N^{2} + 76 N + 156$ \\
37 & $e123|e24|33|5|555||$ &\ 0.0005695(6) &\ 6.5(6)e-06 & 1/6 & $15 N^{2} + 96 N + 132$ \\
38 & $e123|e24|34|55|55||$ &\ 0.0588741(4) & -0.0004036(7) & 1/4 & $N^{3} + 14 N^{2} + 76 N + 152$ \\
39 & $e123|e24|35|45|55||$ &\ 0.05068868(19) &\ 2.386(35)e-05 & 1 & $7 N^{2} + 72 N + 164$  \\
40 & $e123|e24|55|445|5||$ &\ 0.0593005(4) & -0.0004564(7) & 1/4 & $11 N^{2} + 76 N + 156$ \\
41 & $e123|e45|334|5|55||$ &\ 0.0819720(9) &\ 0.0001271(10) & 1/8 & $11 N^{2} + 76 N + 156$ \\
42 & $e123|e45|344|55|5||$ &\ 0.0596861(6) &\ 0.0004084(9) & 1/8 & $N^{3} + 10 N^{2} + 72 N + 160$ \\
43 & $e123|e45|345|45|5||$ &\ 0.04337074(32) &\ 0.00049365(32) & 1/4 & $5 N^{2} + 62 N + 176$ \\
44$^*$ & $e123|e45|444|555|||$ &\ 1.3186(13)e-05 &\ 1.3186(13)e-05 & 1/72 & $27 N^{2} + 108 N + 108$ \\
45 & $e123|e45|445|455|||$ &\ 0.0569110(6) & -0.0001072(8) & 1/8 & $7 N^{2} + 72 N + 164$ \\
46 & $e123|224|4|555|e5||$ & -0.0004155(8) &\ 0.0001011(8) & 1/24 & $15 N^{2} + 96 N + 132$ \\
47 & $e123|224|5|445|5|e|$ &\ 0.0900947(7) &\ 0.0007530(8) & 1/4 & $7 N^{2} + 72 N + 164$ \\
48 & $e123|234|45|45|5|e|$ &\ 0.04973005(25) & -0.01052445(25) & 1/2 & $2 N^{2} + 55 N + 186$ \\
49 & $e123|234|45|55|e5||$ &\ 0.05809943(29) &\ 0.0025797(5) & 1/2 & $5 N^{2} + 62 N + 176$ \\
50 & $e123|245|45|445||e|$ &\ 0.0553498(4) & -0.0001699(6) & 1/4 & $5 N^{2} + 62 N + 176$ \\
\hline\hline 
\caption{The values of the individual self energy diagrams in 6 loops}
\label{table_V}
\end{longtable}
\endgroup
\end{center}